\documentclass[aps,physrev,graphicx,amsmath,amssymb,reprint]{revtex4-2} 
\usepackage{graphicx}
\usepackage{subfig}
\usepackage{physics}
\usepackage{hyperref}
\usepackage{mathtools}

\DeclarePairedDelimiter\floor{\lfloor}{\rfloor}
\usepackage[version=3]{mhchem} 
\usepackage{dcolumn}
\newcolumntype{d}[1]{D{.}{\cdot}{#1} }
\usepackage{bm}
\usepackage{xcolor,soul}
\usepackage{float}
\usepackage{tikz}
\usepackage[font=small,labelfont=bf,
   justification=Justified,
   singlelinecheck=off,
   format=plain]{caption} 
\usetikzlibrary{decorations.markings}

\newcommand*\pFqskip{8mu}
\catcode`,\active
\newcommand*\pFq{\begingroup
        \catcode`\,\active
        \def ,{\mskip\pFqskip\relax}%
        \dopFq
}
\catcode`\,12
\def\dopFq#1#2#3#4#5{%
        {}_{#1}F_{#2}\biggl[\genfrac..{0pt}{}{#3}{#4};#5\biggr]%
        \endgroup
}

\usepackage{mathpazo} 

\usepackage{siunitx}

\begin{document}
\title{
Multiple phonon modes in Feynman path-integral variational polaron mobility
}
\date{\today} 

\author{Bradley A. A. Martin}
\affiliation{Department of Physics, Imperial College London, Exhibition Road, London SW7 2AZ, UK}

\author{Jarvist Moore Frost}
\affiliation{Department of Physics, Imperial College London, Exhibition Road, London SW7 2AZ, UK}
\affiliation{Department of Chemistry, Imperial College London, White City Campus, London W12 0BZ, UK}
\email[Electronic mail:]{jarvist.frost@imperial.ac.uk}


\begin{abstract}
The Feynman path-integral variational approach to the polaron
    problem\cite{Feynman1955}, along with the associated FHIP linear-response
    mobility theory\cite{Feynman1962}, provides a computationally amenable
    method to predict the frequency-resolved temperature-dependent
    charge-carrier mobility, and other experimental observables in polar
    semiconductors. We show that the FHIP mobility theory predicts non-Drude
    transport behaviour, and shows remarkably good agreement with the recent
    diagrammatic Monte-Carlo mobility simulations of Mishchenko et
    al.\cite{Mishchenko2019} for the abstract Fr\"ohlich Hamiltonian.  

    We extend this method to multiple phonon modes in the Fr\"ohlich model
    action.  
    This enables a slightly better variational solution, as inferred from the
    resulting energy.  
    We carry forward this extra complexity into the mobility theory, where it
    shows richer structure in the frequency and temperature dependent mobility,
    due to the different phonon modes activating at different energies.

    The method provides a computationally efficient and fully quantitative
    method of predicting polaron mobility and response in real materials.
\end{abstract}




\maketitle 

\section{Introduction}\label{introduction}

An excess electron in a polar semiconductor polarises and distorts the surrounding lattice. This polarisation then attempts to localise the electron, forming a quasi-particle state known as the polaron. 

When the electron-phonon coupling is large, the extent of the polaron
wavefunction becomes comparable to the lattice constant and a small polaron is
formed where details of the interaction with the atoms are important in
determining polaron properties, but the polaron itself is localised. Many
studies have investigated the properties of small polarons\cite{holstein1959,
lang1963kinetic, lang1964mobility, Emin1993, alexandrov1999mobile, Sio2019,
franchini2021}, where many analytical and numerical studies have primarily
focused on the Holstein model with a short-range electron-phonon
interaction\cite{holstein1959}.

If the competition between the localising potential and the electron kinetic
energy results in a large-polaron state, larger than the unit cell, a continuum
approximation is valid and the details of the interaction with the atoms can be
ignored, but the polaron itself is a dynamic object. 
The most simple large polaron model was introduced by Fr\"ohlich\cite{Frohlich1954}, of a single Fermion (the electron) interacting with an infinite field of Bosons (the phonon excitations of the lattice). 
A major simplification with regards to real materials is assuming that only one phonon mode (which is the longitudinal optical mode, of a binary material) is infrared active (thus having dielectrically mediated electron-phonon interaction), and that this mode is dispersionless. 
The reciprocal-space integrals then become analytic with closed form. 
This Fr\"ohlich model is described by the Hamiltonian

\begin{equation}
    \hat{H} = \frac{\vb{p}^2}{2m^*} + \sum_\mathbf{k} \hbar \, \omega_{0} \, a_\mathbf{k}^\dagger a_\mathbf{k} + \sum_\mathbf{k} ( V_\mathbf{k} \, a_\mathbf{k} \, e^{i\mathbf{k} \cdot \mathbf{r}} + V_\mathbf{k}^* \, a_\mathbf{k}^\dagger \, e^{-i\mathbf{k} \cdot \mathbf{r}}) .
\label{eqn:frohlich}
\end{equation}

Here $\mathbf{r}$ is the electron vector position, $\mathbf{p}$ its conjugate momentum, $m^*$ the electron effective mass, $\hbar$ the reduced Planck constant, $\omega_{0}$ the longitudinal optical phonon frequency, $a_\mathbf{k}^\dagger, a_\mathbf{k}$ the phonon creation and annihilation operators with phonon wavevector $\mathbf{k}$. The electron-phonon coupling parameter is 
\begin{equation}
    V_{\mathbf{k}} = i \frac{2 \hbar\omega_0}{\abs{\mathbf{k}}} \left( \sqrt{\frac{\hbar}{2 m^* \omega_0}} \frac{\alpha \pi}{\Omega_0} \right)^{\frac{1}{2}} .
\end{equation}
Here $\Omega_0$ is the unit cell volume, $\alpha$ is Fr\"ohlich's dimensionless interaction parameter and other variables are as above.
The model is entirely characterised by the unit-less parameter $\alpha$. 

Though this seems highly idealised, the $\alpha$ parameter is a direct function
of the semiconductor properties: an effective mass $m^*$ (modelling the
relevant band-structure of the charge carrier), a phonon frequency $\omega_{0}$
(the quantisation of the phonon field), and the dielectric electron-phonon
coupling between them (which dominates for polar materials\cite{Giustino2017}), 

\begin{equation}
    \alpha = \frac{1}{2} \left( \frac{1}{\epsilon_{\textrm{optic}}}
    - \frac{1}{\epsilon_{\textrm{static}}} \right) \frac{e^2}{\hbar\omega_{0}} \left(
    \frac{2m^*\omega_{0}}{\hbar} \right)^{\frac{1}{2}} .
\label{eqn:alpha}
\end{equation}

Fr\"ohlich's Hamiltonian, though describing a simple physical system of
a single effective mass electron coupled to a single-frequency phonon field,
has resisted exact solution. 
This is a quantum field problem, as the phonon occupation numbers can change. 

One celebrated approximation is Feynman's variational path-integral approach\cite{Feynman1955}. 
This method is surprisingly accurate\cite{Hahn2018} considering the light computational effort, and applies for the full range of the Fr\"ohlich $\alpha$ electron-phonon coupling parameter, without having to make any weak- or strong-coupling approximation. 
The method was extended by Feynman-Hellwarth-Iddings-Platzman\cite{Feynman1962} (commonly referred to as `FHIP') to offer a prediction of temperature dependent mobility (in the linear-response regime) for polar materials, without any empirical parameters, and without resorting to perturbation theory. This method was alternatively derived and used by Peeters and Devreese\cite{Peeters1981, Peeters1983, Peeters1984, Peeters1986}.
The textbook definition of the `FHIP' dc-mobility is an asymptotic
solution recovered from a power series expansion of the model action around a solvable quadratic trial action. The resultant impedance function is well defined and analytic across all frequencies, temperatures and polaron couplings ($\alpha$). 
A generalisation to finite temperatures was made by \=Osaka\cite{Osaka1959},
and the addition of an external driving force by Castrigiano and
Kokiantonis\cite{Castrigiano1983, Castrigiano1984}, and
Saitoh\cite{Saitoh1980}.

Hellwarth and Biaggio\cite{Hellwarth1999} provide a method to replace the
multiple phonon modes of a complex material with a single effective frequency
and coupling. 
This approach has been used by ourselves\cite{Frost2017, FrostJOSS2018,
Zheng2021} and others\cite{Sendner2016} to predict phenomenological properties
of charge-transport for direct comparison to experiment.

In this paper we first describe the Feynman variational quasi-particle polaron
approach\cite{Feynman1955}, providing a consistent description with modern
nomenclature and notation. 
We then show that the FHIP mobility theory\cite{Feynman1962} predicts non-Drude
transport behaviour, and agrees closely with recent diagrammatic Monte-Carlo
mobility predictions of Mishchenko et al.\cite{Mishchenko2019}, for the
abstract Fr\"ohlich Hamiltonian.

Second we extend the method to more accurately model complex real materials by
explicitly including multiple phonon modes. 
Taking a multimodal generalisation of the Fr\=ohlich Hamiltonian we derive
a multimodal version of the Feynman-Jensen variational expression for the free
energy at all temperatures, and then follow the methodology of
FHIP\cite{Feynman1962}, to derive expressions for the temperature- and
frequency-dependent complex impedance. 

Using the example of the well-characterised methylammonium lead-halide
(MAPbI$_3$) perovskite semiconductor, we provide new estimates of dc mobility
and complex conductivity, which can now be directly measured with transient
Terahertz conductivity measurements\cite{Zheng2021}. 

A key technical discovery during this work is that direct numerical integration of the memory function $\chi(\Omega)$ of FHIP\cite{Feynman1962} (required to calculate the polaron mobility), rather than the commonly used contour-rotated integral, has more easily controlled errors for frequency-dependent properties. This is significant as many previous attempts\cite{Feynman1962, Devreese1972, Hellwarth1999, Frost2017} (including ourselves), numerically evaluate the contour-rotated integral using complicated and computationally expensive power-series expansions in terms of special functions. This can be avoided entirely.

As this method requires a relatively trivial amount of computer time, has controlled errors (no Monte-Carlo sampling, or analytic continuation, limitations), and offers
scope for further expansion and refinement, we suggest that it will be useful
to predict polar semiconductor transport properties, particularly in the
computational identification of new semiconductors for renewable energy
applications.

\section{The path integral approach to the Fr\"ohlich polaron} \label{Sec:theory}

\subsection{Feynman variational approach} \label{Sec:feynmanvar}

The 1955 Feynman variational approach\cite{Feynman1955} casts the Fr\"ohlich
polaron problem into a Lagrangian path and field integral (the \emph{model}
action), and then integrates out the infinite quantum field of phonon
excitations. The result is a remapping to an effective quasi-particle
Lagrangian path integral, where an electron is coupled by a non-local two-time
Coulomb potential to another fictitious massive particle, representing the
disturbance in the lattice generated by its passage at a previous time. 
The density matrix $\rho$ for the electron to go from position $\mathbf{r'}$ to $\mathbf{r''}$ within an imaginary time $i\hbar\beta$ is
\begin{equation}
    \rho(\mathbf{r'}, \mathbf{r''}; \hbar\beta) = \int_{\mathbf{r}(0) = \mathbf{r'}}^{\mathbf{r}(\hbar\beta) = \mathbf{r''}} \mathcal{D}\mathbf{r}(\tau) \exp{\left(-\frac{S[\mathbf{r}(\tau)]}{\hbar}\right)}
.
\end{equation}
The \emph{model} action  $S$ for the Fr\"ohlich polaron is
\begin{equation} \label{eqn:polaron_action}
    \begin{aligned}  S[\mathbf{r}(\tau)] &= \frac{m^*}{2}\int^{\hbar\beta}_0
        d\tau \left(\frac{d\mathbf{r}(\tau)}{d\tau}\right)^2
        -  \frac{(\hbar\omega_0)^{\frac{3}{2}} \alpha}{2 \sqrt{2 m^*}} \\
         & \quad \times \int^{\hbar\beta}_0 d\tau \int^{\hbar\beta}_0 d\sigma\  \frac{g_{\omega_0}(|\tau - \sigma|)}{|\mathbf{r}(\tau) - \mathbf{r}(\sigma)|}
    \end{aligned} ,
\end{equation} 
and where 
\begin{equation}
    g_{\omega_0}(\tau) = \frac{\cosh(\omega_0 (\tau - \hbar \beta / 2))}{\sinh(\hbar \omega_0 \beta / 2)}
\end{equation}
is the \emph{imaginary-time} phonon correlation function. 

We cannot easily evaluate the path-integral for the $1/r$ Coulomb potential, so Jensen's inequality, $\langle \exp{f} \rangle \geq \exp{\langle f \rangle}$, is used to approximate the effective Lagrangian by an analytically path-integrable non-local two-time \emph{quadratic} Lagrangian (the \emph{trial} action), $S_0$, 

\begin{equation} \label{eqn:trial_action}
    \begin{aligned}
        S_0[\mathbf{r}(\tau)] &= \frac{m^*}{2}\int^{\hbar\beta}_0 d\tau \left(\frac{d\mathbf{r}(\tau)}{d\tau}\right)^2 + \frac{C}{2} \int^{\hbar\beta}_0 d\tau \\
        &\quad\times \int^{\hbar\beta}_0 d\sigma\ g_{w \omega_0}(|\tau - \sigma|) \left(\mathbf{r}(\tau) - \mathbf{r}(\sigma)\right)^2 .
    \end{aligned}
\end{equation}
The resulting Feynman-Jensen inequality gives a solvable upper-bound to the
(model) free energy,
\begin{equation}\label{eqn:feynman_jensen}
    F \leq F_0 + \langle S - S_0 \rangle_0 ,
\end{equation}
where $F_0$ is the free energy of the trial system and $\langle S - S_0
\rangle_0$ is the expectant difference in the two actions, evaluated with respect to the trial system,
\begin{equation}
    \langle S-S_0 \rangle_0 = \frac{\int \mathcal{D}\textbf{r}(\tau) (S-S_0) e^{-S_0[\textbf{r}]/ \hbar}}{\int \mathcal{D}\textbf{r}(\tau) e^{-S_0[\textbf{r}]/\hbar}}.
\end{equation}
The process is variational, in that the $C$ (a harmonic coupling term) and $w$
(which controls the exponential decay rate of the interaction in
imaginary-time) parameters are varied to minimise the RHS of Eqn.
(\ref{eqn:feynman_jensen}), giving the lowest upper-bound to the free energy. 
Diagramatic Monte-Carlo shows that this method approches the true energy across a wide range of coupling parameters\cite{Mishchenko2000,Hahn2018,Mishchenko2019}. 

\subsection{The FHIP mobility} \label{Sec:FHIP}

Feynman-Hellwarth-Iddings-Platzman\cite{Feynman1962} (FHIP) derive an expression for the linear response of the Fr\"ohlich polaron to a weak, spatially uniform, time-varying electric field $\textbf{E}(t) = E_0 \exp(i\Omega t)$, where $\Omega$ is the angular  frequency of the field. The field induces a current due to the movement of the electron,
\begin{equation}
    \textbf{j}(\Omega) = \frac{\textbf{E}(\Omega)}{z(\Omega)} = e \frac{d}{dt}\langle \mathbf{r}(t) \rangle ,
\end{equation}
where $z(\Omega)$ is the complex impedance function and $\langle \mathbf{r}(t) \rangle$ the expectation of the electron position. 
For sufficiently weak fields (in the linear response regime), the relationship between the impedance and the Fourier transform of the Green's function $G(t)$ of the polaron is
\begin{equation}
    \int_{-\infty}^{\infty} dt G(t)\ e^{-i \Omega t} = G(\Omega) = \frac{1}{\Omega z(\Omega)},
\end{equation}
where $G(t) = 0$ for $t < 0$. 

The electric field $E(t)$ appears as an addition linear term in the Fr\"ohlich Hamiltonian, $-\mathbf{E} \cdot \mathbf{r}$. The expected electron position can be evaluated from the density matrix $\rho(t)$ of the system,
\begin{equation}
    \langle \mathbf{r}(t) \rangle = \mathrm{Tr}\left\{ \mathbf{r} \rho(t) \right\}.
\end{equation}
Assuming that the system is initially in thermal equilibrium $\rho_0 = \exp\left(-\beta H\right)$, the time evolution of the density matrix is evaluated with
\begin{equation}
    i\hbar \frac{\partial \rho}{\partial t} = \left[ H, \rho\right].
\end{equation}
Therefore, the density matrix at some later time $t$ is
\begin{equation}
    \rho(t) = U(t) \rho_0 U'^\dagger(t),
\end{equation}
where the unitary operators $U$ and $U'$ for time evolution are
\begin{equation}
    \begin{aligned}
        U(t) &= \exp \left\{ -\frac{i}{\hbar} \int^t_0 \left[ H(s)
        - \mathbf{r}(s) \cdot \mathbf{E}(s) \right] ds \right\} , \\
        U'(t) &= \exp \left\{ -\frac{i}{\hbar} \int^t_0 \left[ H'(s) - \mathbf{r'}(s) \cdot \mathbf{E'}(s) \right] ds \right\} .
    \end{aligned}
\end{equation}
Here unprimed operators are time-ordered with latest times on the far left, whereas primed operators are oppositely time-ordered with latest times on the far right. (Technically the electric field $E(t)$ is not an operator in need of time-ordering, but it is useful treat $E$ and $E'$ as different arbitrary functions.)

FHIP assumes that the initial state is a product state of the phonon bath and
the electron system, where only the phonon oscillators are initially in thermal equilibrium, $\rho_0 \propto \exp\left(-\hbar\beta \sum_{\mathbf{k}} \omega_{\mathbf{k}} b^\dagger_{\mathbf{k}} b_{\mathbf{k}}\right)$ at temperature $T = (k_B\beta)^{-1}$. 
The true system would quickly thermalise to the temperature of the (much larger) phonon
bath, but the linear Feynman polaron model cannot since it is entirely
harmonic. 
As shown by Sels\cite{DriesSelsThesis}, it would be more correct to impose that
the entire model system starts in thermal equilibrium.
This error results in the lack of a `$2\beta$' dependence in FHIP 
(low-temperature) dc-mobility. 

We can formulate $\mathrm{Tr}\{\rho(t)\}$ as a path integral generating functional
\begin{equation}
    \begin{aligned}
        \mathrm{Tr}\{\rho(t)\} &\equiv \mathcal{Z}\left[ \mathbf{E}(t), \mathbf{E'}(t) \right] \\
        &= \int \mathcal{D}\mathbf{r}(t) \mathcal{D}\mathbf{r'}(t) \exp \left\{ \frac{i}{\hbar} \Phi\left[ \mathbf{r}(t), \mathbf{r'}(t) \right] \right. \\
        &\left.\quad + \frac{i}{\hbar} \left( S[\mathbf{r}(t), \mathbf{E}(t)] - S[\mathbf{r'}(t), \mathbf{E'}(t)] \right) \right\},
    \end{aligned}
\end{equation}
where $S$ is the classical action of the uncoupled electron,
\begin{equation}
    \begin{aligned}
        S[\mathbf{r}(t), \mathbf{E}(t)] = \int_0^t ds\ \left[ \frac{m^*}{2} \left(\frac{d\mathbf{r}(s)}{ds}\right)^2 + \mathbf{E}(s) \cdot \mathbf{r}(s) \right].
    \end{aligned}
\end{equation}
$\Phi\left[\mathbf{r}(t), \mathbf{r'}(t)\right]$ is the phase of the \emph{influence functional}\cite{FeynmanVernon1963}. 
The influence functional phase for the Fr\"ohlich model $\Phi_F\left[\mathbf{r}(t), \mathbf{r'}(t)\right]$ is derived from the model action (Eqn. (\ref{eqn:polaron_action})) and is given by
\begin{equation}
    \begin{aligned}
          \Phi_F\left[\mathbf{r}(t), \mathbf{r'}(t)\right] &=
          \frac{i(\hbar\omega_0)^{\frac{3}{2}}\alpha}{2\sqrt{2 m^*}} \int^\infty_{-\infty} dt \int^\infty_{-\infty} ds\ \left[ \frac{g_{\omega_0}\left(|t-s|\right)}{|\mathbf{r}(t) - \mathbf{r}(s)|} \right. \\
          &\left.\quad + \frac{g_{\omega_0}^*(|t-s|)}{|\mathbf{r'}(t) - \mathbf{r'}(s)|} - 2\frac{g_{\omega_0}(t-s)}{|\mathbf{r'}(t) - \mathbf{r}(s)|} \right] , 
    \end{aligned}
\end{equation}
where $g_{\omega_0}(t)$ is the \emph{real-time} phonon Green's function and is given by
\begin{equation}
    g_{\omega_0}(t) = \frac{\cos\left(\omega_0(t - i\hbar\beta/2)\right)}{\sinh\left(\hbar\omega_0\beta/2\right)}.
\end{equation}
The double path integral is over closed paths satifying the boundary condition $\mathbf{r}(t) - \mathbf{r'}(t) = 0$ as $t \rightarrow \pm \infty$. \newline

The Green's function $G(t - t')$ is the response to a $\delta$-function electric field $\mathbf{E}(s) = \bm{\epsilon} \delta(s - t) = \mathbf{E'}(s)$. It can be evaluated from the first functional derivative of the generating functional $\mathcal{Z}[\mathbf{E}(t), \mathbf{E'}(t)]$ with respect to $\mathbf{E}(t) - \mathbf{E'}(t)$. We can formulate the primed and unprimed electric fields as
\begin{equation}
    \begin{aligned}
          \mathbf{E}(s) &= \bm{\epsilon} \delta(s - t) + \bm{\eta} \delta(s - t') \\
          \mathbf{E'}(s) &= \bm{\epsilon} \delta(s - t) - \bm{\eta} \delta(s - t'). 
    \end{aligned}
\end{equation}
This reduces the generating \textit{functional} $\mathcal{Z}[\mathbf{E}(t), \mathbf{E'}(t)]$ into a generating \textit{function} $\mathcal{Z}(\bm{\epsilon}, \bm{\eta})$. The Green's function can then be evaluated from
\begin{equation}
    \begin{gathered}
    G(t - t') = -\frac{\hbar^2}{2} \frac{1}{\mathcal{Z}(0, 0)} \frac{\partial^2 \mathcal{Z}(\bm{\epsilon}, \bm{\eta})}{\partial \bm{\epsilon} \partial \bm{\eta}} \Bigg|_{\bm{\epsilon} = \bm{\eta} = 0}.
    \end{gathered}
\end{equation}

In FHIP the generating function $\mathcal{Z}(\bm{\epsilon},
\bm{\eta})$ is approximated by taking the zeroth ($\mathcal{Z}_0$) and first order
($\mathcal{Z}_1$) terms from an expansion of the path integral around an
exactly solvable harmonic system. 
This system is described by a quadratic influence functional phase
$\Phi_0\left[\mathbf{r}(t), \mathbf{r'}(t)\right]$ that has been derived from
the quadratic trial action $S_0$ (Eqn.
(\ref{eqn:trial_action}) ) and is given by \begin{equation}
    \begin{aligned}
        \Phi_0\left[\mathbf{r}(t), \mathbf{r'}(t)\right] &= -\frac{iC}{2} \int^\infty_{-\infty} dt \int^\infty_{-\infty} ds\ \left[ \frac{g_{w \omega_0}\left(|t-s|\right)}{|\mathbf{r}(t) - \mathbf{r}(s)|^{-2}} \right. \\
          &\left.\quad + \frac{g_{w \omega_0}^*(|t-s|)}{|\mathbf{r'}(t) - \mathbf{r'}(s)|^{-2}} - 2\frac{g_{w \omega_0}(t-s)}{|\mathbf{r'}(t) - \mathbf{r}(s)|^{-2}} \right] ,
    \end{aligned}
\end{equation}
where $C$ and $w$ are Feynman's variational parameters. 
$\mathcal{Z}(\bm{\epsilon}, \bm{\eta})$ is then approximated by two terms, 
\begin{equation}
    \begin{aligned}
    \mathcal{Z}(\bm{\epsilon}, \bm{\eta}) &= \int \mathcal{D}\mathbf{r} \mathcal{D}\mathbf{r'}\ e^{\frac{i}{\hbar} (S[\mathbf{r}] - S[\mathbf{r'}] + \Phi[\mathbf{r}, \mathbf{r'}])} \\
    &\approx \int \mathcal{D}\mathbf{r} \mathcal{D}\mathbf{r'}\ e^{\frac{i}{\hbar} (S[\mathbf{r}] - S[\mathbf{r'}] + \Phi_0[\mathbf{r}, \mathbf{r'}])} \\
    &\quad\times \left[ 1 +  \frac{i}{\hbar}\left(\Phi[\mathbf{r}, \mathbf{r'}] - \Phi_0[\mathbf{r}, \mathbf{r'}]\right)\right] \\
    &\equiv \mathcal{Z}_0 + \mathcal{Z}_1 .
    \end{aligned}
\end{equation}

In FHIP and Devreese et al.\cite{Devreese1972} they find that it is more accurate to use the complex impedance function over the complex conductivity $\sigma(\Omega)$ ($=1 / z(\Omega)$) by taking the Taylor expansion of the impedance,

\begin{equation}
    \begin{aligned}
    \Omega z(\Omega) = \frac{1}{G(\Omega)} & \approx \frac{1}{G_0(\Omega) + G_1(\Omega)} \\
    &\approx \frac{1}{G_0(\Omega)} - \frac{1}{G_0^2(\Omega)} G_1(\Omega) ,
    \end{aligned}
\end{equation}
where $G_0$ and $G_1$ are the classical and first-order quantum correction response functions obtained from $\mathcal{Z}_0$ and $\mathcal{Z}_1$ respectively. 

This expansion of the impedance gives
\begin{equation}\label{eqn:compleximpedance}
    z(\Omega) \approx i\left(\Omega - \frac{\chi(\Omega)}{\Omega}\right) ,
\end{equation}
where
\begin{equation}\label{eqn:fhip_chi}
    \chi(\Omega) = \frac{2 \alpha \omega_{0}^{2}}{3 \sqrt{\pi}} \int_0^\infty dt\ \left( 1 - e^{i \Omega t} \right) \Im{S(t)} ,
\end{equation}
is a memory function that contains all the first-order corrections from the electron-phonon interactions (Eqn. (35a) in FHIP). 

Here $S(t)$ (Eqn. (35b) in FHIP) is proportional to the dynamic structure factor for the electron and is given by
\begin{equation}
    S(t) = g_{\omega_0}(t) \left[D(t)\right]^{-\frac{3}{2}} 
\end{equation}
where
\begin{equation}\label{eqn:D}
    \begin{aligned}
    D(t) &= 2 \frac{v^2-w^2}{v^3} \frac{\sin(v\omega_0t/2) \sin(v\omega_0[t-i\hbar\beta])}{\sinh(v\omega_0\hbar\beta/2)}\\
    &\quad-i\frac{w^2}{v^2}\omega_0t\left(1-\frac{t}{i\hbar\beta}\right).
    \end{aligned}
\end{equation}
Our $D(t)$ is the same as $D(u)$ in Eqn. (35c) in FHIP. The frequency-dependent mobility $\mu(\Omega)$ is obtained from the impedence by using
\begin{equation}\label{eqn:freq_dep_mobility}
\begin{aligned}
    \mu(\Omega) &= \textrm{Re}\left\{\frac{1}{z(\Omega)}\right\}\\
    &= \frac{e}{m^*} \frac{\Omega\ \textrm{Im}\chi(\Omega)}{\Omega^4 - 2\ \Omega^2\  \textrm{Re}\chi(\Omega) + |\chi(\Omega)|^2} ,
    \end{aligned}
\end{equation}
where the values of the variational parameters $v$ and $w$ are those that minimise the polaron free energy in Eqn. (\ref{eqn:feynman_jensen}). In the limit that the frequency $\Omega \rightarrow 0$ gives the FHIP dc-mobility,
\begin{equation}\label{eqn:dcmobility}
    \mu^{-1}_{dc} = \frac{m^*}{e} \lim_{\Omega \rightarrow 0} \frac{\textrm{Im}\chi(\Omega)}{\Omega} ,
\end{equation}
since $\textrm{Re}\chi(\Omega = 0) = 0$.

\subsection{Numerical evaluation of the memory function} \label{Sec:memoryfuncnum}

In summary, the integral for the memory function is 

\begin{subequations}\label{eqn:memoryfunction}
\begin{align}
    \chi(\Omega) &= \frac{2 \alpha \omega_{0}^{2}}{3 \sqrt{\pi}} \int^\infty_0 dt \left( 1 - e^{i\Omega t} \right) \Im{S(t)} \\
    S(t) &= \frac{\cos(\omega_0 (t
    - i\hbar\beta/2))}{\sinh(\hbar\omega_0\beta/2)}
    \left[D(t)\right]^{-\frac{3}{2}} \\
    D(t) &= 2\frac{v^2 - w^2}{v^3} \frac{\sin(v\omega_0 t / 2) \sin(v \omega_0 (t - i\hbar\beta))}{\sinh(v \omega_0 \hbar \beta / 2)} \notag\\
    &\quad- i \frac{w^2}{v^2} \omega_0 t \left(1 - \frac{t}{i\hbar\beta}\right),
\end{align}
\end{subequations}
where $\Omega$ is the angular frequency of the driving electric field, $\omega_0$ is the angular phonon frequency, $\beta = 1/k_B T$ is the thermodynamic temperature and $v$ and $w$ are variational parameters whose values minimise the polaron free energy. 
This is the same as Eqns. (35) in FHIP, but in SI units and with an alternative algebra. 

Previous work\cite{Feynman1962, Devreese1972}, including our own (see Appendices), made use of the `doubly-oscillatory' contour-rotated integral for the complex memory function in Eqn. (\ref{eqn:fhip_chi}). The imaginary component of the memory function is (Eqns. (47) in Ref. \onlinecite{Feynman1962}),
\begin{widetext}
\begin{equation}\label{eqn:powerseriesimag}
        \Im{\chi(\Omega)} = \frac{2\alpha \omega_0^2}{3\sqrt{\pi}}
        \frac{(\hbar\omega_0\beta)^{\frac{3}{2}} \sinh(\hbar \Omega \beta
        / 2)}{\sinh(\hbar \omega_0 \beta / 2)} \left(\frac{v}{w}\right)^3
        \int_0^\infty d\tau \frac{\cos(v \omega_0 \tau) \cos(\omega_0
        \tau)}{\left[ \omega_0^2 \tau^2 + a^2 - b \cos(v \omega_0 \tau)
        \right]^{\frac{3}{2}}},
\end{equation}
\end{widetext}
where $a^2 \equiv\left(\hbar\omega_0\beta/2\right)^2 + R \hbar\beta\omega_0 \coth(\hbar \beta \omega_0 v/2)$, $b \equiv R \hbar\beta\omega_0 / \sinh(\hbar\beta\omega_0 v / 2)$ and $R \equiv (v^2 - w^2) / (w^2 v)$, and where $\tau$ labels \textit{imaginary time} compared to $t$ that labels \textit{real} time in Eqns. (\ref{eqn:fhip_chi}). Additionally, the contour integral for the real component of the memory function, derived by us (see Appendix A), is,
\begin{widetext}
    \begin{equation}\label{eqn:powerseriesreal}
        \begin{aligned}
            \Re{\chi(\Omega)} = \frac{2\alpha\omega_0^2}{3\sqrt{\pi}}
            \frac{(\hbar\omega_0\beta)^{\frac{3}{2}}}{\sinh(\hbar\omega_0\beta/2)}\left(\frac{v}{w}\right)^3
            &\Biggl\{\sinh\left(\frac{\hbar \Omega
            \beta}{2}\right)\int_0^\infty d\tau \frac{\sin(\Omega \tau)
            \cos(\omega_0 \tau)}{\left[\omega_0\tau^2 + a^2 - b\cos(v \omega_0
            \tau)\right]^{\frac{3}{2}}} \\
            &\quad- \int^{\frac{\hbar\beta}{2}}_0 d\tau
            \frac{(1-\cosh(\Omega(\tau - \hbar\beta/2))
            \cosh(\omega_0\tau)}{\left[a^2 - \omega_0^2 \tau^2
            - b\cosh\left(v\omega_0\tau\right) \right]^{\frac{3}{2}}} \Biggr\}.
        \end{aligned}
    \end{equation}
\end{widetext}
The imaginary component of the memory function can be expanded in Bessel functions (originally the derivation was outlined in Refs.\cite{Feynman1955, Devreese1972}, but in Appendix B we provide an in-depth derivation) and the real component in terms of Bessel and Struve functions (see Appendix C for a derivation of the expansion that we believe to be new).

However, we found that the cost of evaluating these expansions became large at low temperatures, requiring use of arbitrary-precision numerics to slowly reach converged solutions. In Devreese et al.\cite{Devreese1972}, they found an alternative analytic expansion for the real component, but similarly found it to have poor convergence for all temperatures, opting instead to transform the integrand to one that has better convergence.

Instead of using any of the contour integrals or power-series expansions, we found that directly numerically integrating Eqn. (\ref{eqn:memoryfunction}) using an adaptive Gauss-Kronrod quadrature algorithm leads to faster convergence and controlled errors. 
Asymptotic limits of these contour integral expansions, especially at low temperature, may still prove useful.

\section{``Beyond Quasiparticle'' polaron mobility} \label{Sec:beyondquasi}

\begin{figure}[!tbp]
\centering
    \includegraphics[width=\columnwidth]{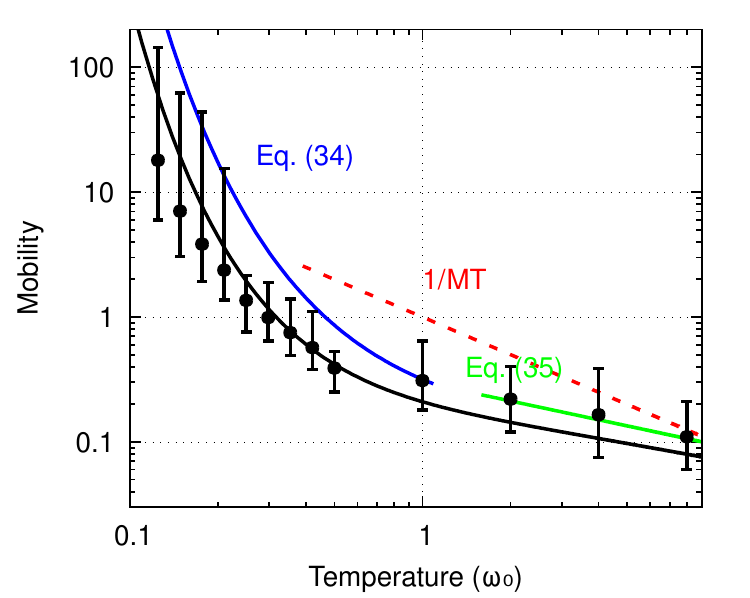}
    \caption{\label{fig:mishchenko_fig2} 
    The FHIP temperature-dependent mobility for the intermediate coupling regime $\alpha = 2.5$ (black, solid), as comparison to Mishchenko et al.\cite{Mishchenko2019} Fig. 2  (black dots, with Monte-Carlo sampling error bars). 
    This electron-phonon coupling strength is most relevant for moderately polar semiconductors. 
    Following Mishchenko, the blue solid line shows the anti-adiabatic / weak-coupling limit of the mobility provided by Eqn. (\ref{eqn:coldmobility}) (Eqn. (5) in\cite{Mishchenko2019}). 
    The green solid line shows the adiabatic limit of the mobility provided by Eqn. (\ref{eqn:hotmobility}) (Eqn. (6) in\cite{Mishchenko2019}). 
    The red dashed line shows the MIR criterion.
    The FHIP method shows good agreement with the limiting behaviour, and is within the Monte-Carlo sampling error of Mishchenko et al.\cite{Mishchenko2019}.
    Already at these relatively weak couplings, the true mobility is well below the MIR independent-scattering criterion. 
    This calls into question the use of the Boltzmann transport equation in simulating even moderately polar materials. 
}
\end{figure}
\begin{figure}[!tbp]
\centering
    \includegraphics[width=\columnwidth]{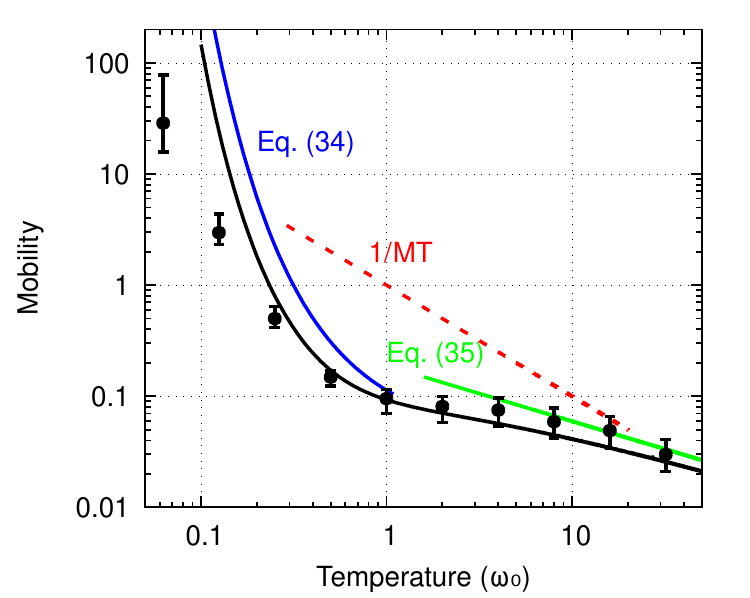}
    \caption{\label{fig:mishchenko_alpha_4} 
    The FHIP temperature-dependent mobility for the intermediate coupling regime $\alpha = 4.0$ (black, solid), as comparison to Mishchenko et al.\cite{Mishchenko2019} Fig. 2 (black dots, with Monte-Carlo sampling error bars). 
    This electron-phonon coupling strength is most relevant for strongly polar semiconductors. 
    Following Mishchenko, the blue solid line shows the anti-adiabatic / weak-coupling limit of the mobility provided by Eqn. (\ref{eqn:coldmobility}) (Eqn. (5) in\cite{Mishchenko2019}). 
    The green solid line shows the adiabatic limit of the mobility provided by Eqn. (\ref{eqn:hotmobility}) (Eqn. (6) in\cite{Mishchenko2019}). 
    The red dashed line shows the MIR criterion.
    The FHIP mobility shows good agreement with the limiting adiabatic and anti-adiabatic behaviour, and is clearly away from the MIR criterion where the Boltzmann transport equation is valid. 
}
\end{figure}
\begin{figure}[!tbp]
\centering
    \includegraphics[width=\columnwidth]{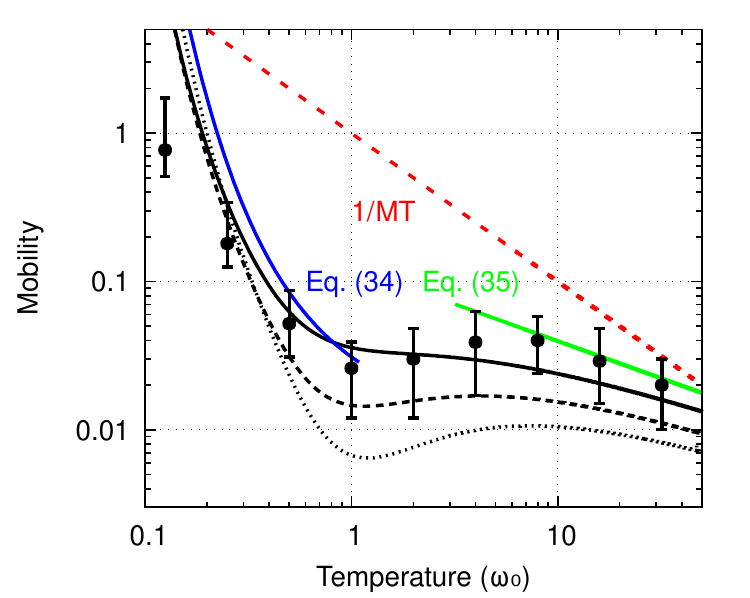}
    \caption{\label{fig:mishchenko_fig3} 
    The FHIP temperature-dependent mobility in the strong-coupling limit, presented as comparison to Mishchenko et al.\cite{Mishchenko2019} Fig. 3 (black dots, with Monte-Carlo sampling error bars).
    Presented are $\alpha = 6$ (black, solid), $\alpha = 8$ (black, dashed), $\alpha = 10$ (black, dotted). 
    The diagrammatic Monte-Carlo results (black, dots, with Monte-Carlo sampling error bars) show pronounced non-monotonic behaviour (the 'hump' at $k_BT/\hbar\omega_0 = 8$) already with $\alpha = 6$, while the FHIP model requires a stronger coupling, though we note that the $\alpha = 6$ FHIP result is within the Monte-Carlo error bars. 
}
\end{figure}

\begin{figure}[!tbp]
    \includegraphics[width=\columnwidth]{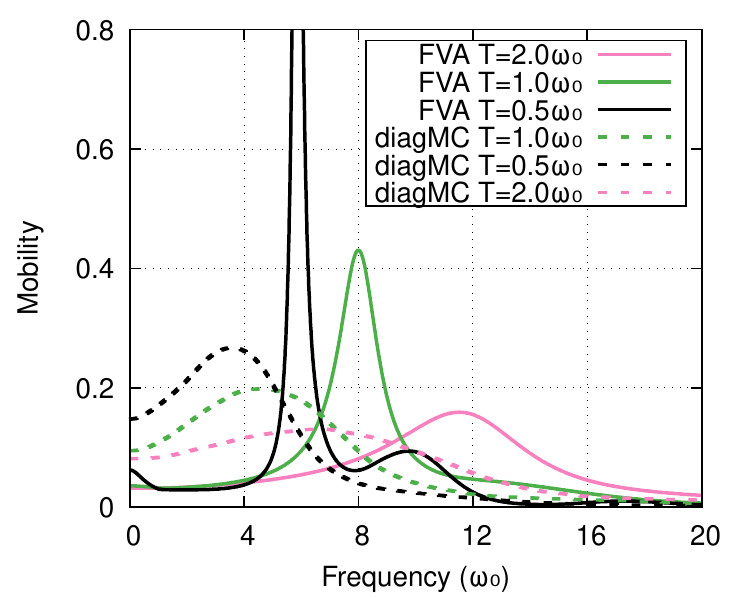}
    \caption{\label{fig:mishchenko_fig4} The FHIP (solid)  frequency- and temperature-dependent mobility, presented as comparison to Mischenko et al.\cite{Mishchenko2019} (dashed). Presented is $\alpha = 6$ for temperatures $k_BT / \hbar\omega_0 = 0.5$ (black), $1.0$ (green) and $2.0$ (pink).
}
\end{figure}

\begin{figure}[!tbp]
    \includegraphics[width=\columnwidth]{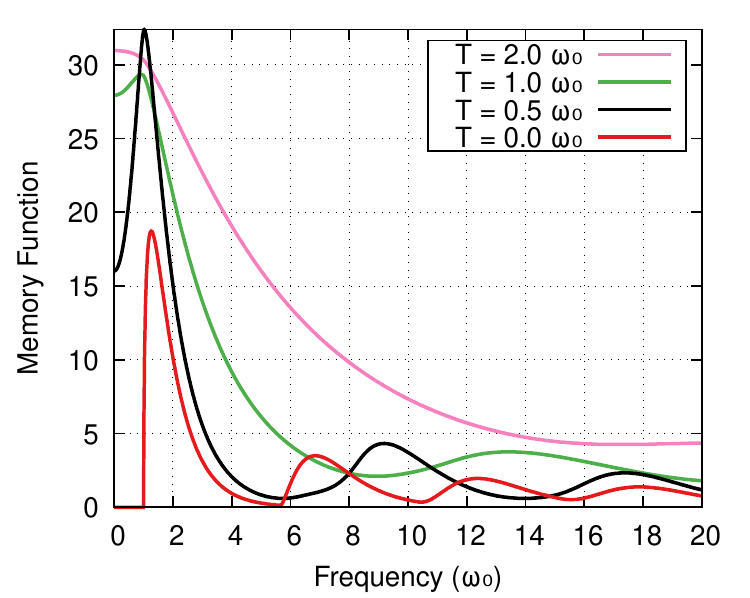}
    \caption{\label{fig:memory_function} The frequency and temperature dependence of the memory function $\chi$ (Eqn. (\ref{eqn:fhip_chi})). The peaks correspond to the Frank-Condon states of the polaron. For zero temperature $k_BT / \hbar\omega_0 = 0.0$ (red) the variational parameter $v = 4.67$ such that the peaks occur after $\omega_0 + nv\omega_0 = \omega_0$, $5.67 \omega_0$, $10.33 \omega_0$ etc. At higher temperatures, the peaks shift to higher frequencies due to the temperature dependence of $v$ that minimises the polaron free energy at a given temperature.
}
\end{figure}

Mishchenko et al.\cite{Mishchenko2019} recently used diagrammatic Monte Carlo (diagMC) calculations to investigate the violation of the so-called ``thermal'' analogue to the Mott-Ioffe-Regel (MIR) criterion in the Fr\"ohlich polaron model. 
This ``thermal'' MIR criterion is perhaps better referred to as the \emph{Planckian bound}\cite{Hartnoll2021} under which a quasiparticle is stable to inelastic scattering. 
For the quasiparticle to propagate coherently, the inelastic scattering time $\tau_{\text{inel}}$ must be greater than the ``Planckian time'' $\tau_{\text{Pl}} = \hbar / k_B T$. 
For the polaron mobility $\mu$, this requires $\mu \gtrsim \frac{e\hbar}{M k_B T}$. 
This Planckian bound can be reformulated into Mishchenko's\cite{Mishchenko2019} ``thermal'' MIR criterion for the validity of the Boltzmann kinetic equation, $l >> \lambda$ where $l$ is the mean free path, and $\lambda$ the de Broglie wavelength, of the charge carrier.

The anti-adiabatic limit ($k_BT \ll \hbar\omega_0$) corresponds to the weak-coupling limit ($\alpha \ll 1$), where the perturbative theory result for the mobility is (Eqn. (5) in Ref. \onlinecite{Mishchenko2019})
\begin{equation}\label{eqn:coldmobility}
    \begin{aligned}
        \mu &= \frac{e}{2M\alpha\omega_0} e^{\hbar \omega_0/k_BT} \\ 
        &= \frac{e}{2m^*\omega_0} \left(\frac{1}{\alpha} - \frac{1}{6}\right) e^{\hbar \omega_0/k_BT}, \quad (k_BT \ll \hbar\omega_0, \alpha \ll 1),
    \end{aligned}
\end{equation}
where $M = m^* / (1 - \alpha / 6)$ is the effective mass renormalisation of the polaron. In the adiabatic regime ($k_BT \gg \hbar\omega_0$), the mobility is obtained from the kinetic equation as (Eqn. (6) in Ref. \onlinecite{Mishchenko2019})
\begin{equation}\label{eqn:hotmobility}
    \mu = \frac{4 e \sqrt{\hbar}}{3\sqrt{\pi}\alpha M \sqrt{\omega_0 k_B T}}, \quad (k_BT \gg \hbar\omega_0),
\end{equation}
which is valid even when $\alpha$ is not small.

Fig. \ref{fig:mishchenko_fig2} is a comparison with Fig. 2 in Mishchenko et al.\cite{Mishchenko2019} of the polaron mobility at $\alpha = 2.5$. At low temperatures ($k_BT \lesssim \hbar\omega_0 / 2$), the exponential behaviour matches the low-temperature mobility in Eqn. (\ref{eqn:coldmobility}). As in\cite{Mishchenko2019}, there appears to be a delay in the onset of the exponential behaviour for $k_BT < \hbar\omega_0$. Likewise, the MIR criterion is violated over the temperature range $0.2 < k_BT/\hbar\omega_0 < 10$. At high temperatures, the FHIP mobility (Eqn. (\ref{eqn:freq_dep_mobility})) has the same $1/\sqrt{T}$ dependence as Eqn. (\ref{eqn:hotmobility}).

In Fig. \ref{fig:mishchenko_fig3} we compare the temperature dependence of the FHIP polaron mobility with the diagMC polaron mobility (Fig. 3 in Mishchenko et al.\cite{Mishchenko2019}) at $\alpha = 6$. The diagMC polaron mobility exhibits non-monotonic behaviour at $\alpha = 6$, with a clear local minimum around $k_BT = \hbar\omega_0$. Here we see similar non-monotonic behaviour in the FHIP mobility with a small local minimum appearing around $k_BT = \hbar\omega_0$ too. However, compared to the diagMC mobility, the local minimum of the FHIP mobility is shallower. The onset of this minimum in the FHIP mobility begins around $\alpha = 6$, with the minimum deepening at stronger couplings ($\alpha = 8,\ 10\ \&\ 12$). Similar to the diagMC mobility, the high-temperature limit is recovered after a maximum at $k_BT/\hbar\omega_0 \sim \alpha$ which shifts with larger $\alpha$. The minimum too appears to be $\alpha$-dependent, occurring at $k_BT/\hbar\omega_0 \sim 1$ for $\alpha = 6$ or $k_BT/\hbar\omega_0 \sim 1.5$ for $\alpha = 12$.

In Fig. \ref{fig:mishchenko_fig4} we compare the temperature and frequency dependence of the FHIP polaron mobility with the diagMC polaron mobility (Fig. 4 in Mishchenko et al.\cite{Mishchenko2019}) at $\alpha = 6$ for temperatures $T = 0.5\omega_0, 1.0\omega_0, 2.0\omega_0$. The FHIP mobility, obtained by integrating Eqn. (\ref{eqn:freq_dep_mobility}), has similar temperature dependence to the diagMC mobility but differs in the frequency response. 

The FHIP mobility shows extra peaks where the first peak is blue-shifted compared to the diagMCs single peak. In\cite{Devreese1972, DeFilippis2006} it is shown that these extra peaks of the FHIP mobility correspond to internal relaxed excited states of the polaron quasiparticle. These internal states correspond to multiple phonon scattering processes. For $k_BT/\hbar\omega_0 = 0.5$, the first peak around $\Omega/\omega_0 \sim 6$ corresponds to one-phonon processes, the peak at $\Omega / \omega_0 \sim 10$ corresponds to two-phonon processes, and so on. This is more clearly seen by analysing the memory function $\chi(\Omega)$ (Eqn. (\ref{eqn:fhip_chi})) at zero temperature, which similarly has peaks at $\Omega / \omega_0 = 1 + nv$, where $n = 0, 1, 2, ...$ and $v$ is one of the Feynman variational parameters (c.f. Fig. \ref{fig:memory_function}). These peaks in the memory function correspond to the same Frank-Condon states. As the temperature increases, the first few peaks become more prominent and broaden due to an increased effective electron-phonon interaction. Eventually, the excitations can no longer be resolved at high temperatures. 

The Feynman variational model of the electron harmonically coupled to a fictitious massive particle (c.f. Section \ref{section:spectra}) lacks a dissipative mechanism for the polaron such that the polaron state described by this model does not lose energy and has an infinite lifetime. However, in de Filippis et al.\cite{DeFilippis2006}, dissipation is included in this model at zero temperature. This attenuates and spreads the harmonic peaks, obscuring the internal polaron transitions, giving closer agreement to the diagMC mobility at zero temperature. We have not used these methods here but they will be investigated in future work to compliment the multiple phonon model action with a more generalised trial action.

\section{Extending the Fr\"ohlich model} \label{Sec:extending}

\subsection{Multiple phonon mode electron phonon coupling} \label{Sec:multicoupling}

In simple cubic polar materials with two atoms in the unit cell, the single triply-degenerate optical phonon branch is split by dielectric coupling into the singly-degenerate longitudinal-optical (LO) mode and double-generate transverse-optical (TO) modes. Only the longitudinal-optical mode is infrared active, and contributes to the Fr\"ohlich dielectric electron-phonon interaction. 

The infrared activity of this mode drives the formation of the polaron. 
Much of the original literature therefore just refers to \emph{the} LO mode. 
In a more complex material the full range of infrared active modes all contribute to the polaron stabilisation, and the infrared activity of these modes is no longer severely restricted by group theory, but are instead best evaluated numerically. 
The driving force of the infrared activity is, however, slightly obscured by the algebra in Eqn. (\ref{eqn:alpha}), and instead this electron-phonon coupling seems to emerge from bulk properties of the lattice. 
The Pekar factor, $\frac{1}{\epsilon_{\textrm{optic}}}-\frac{1}{\epsilon_{\textrm{static}}}$, being particularly opaque.

Rearranging the Pekar factor as
\begin{equation}
    \left( \frac{1}{\epsilon_{\textrm{optic}}} - \frac{1}{\epsilon_{\textrm{static}}} \right) = \frac{\epsilon_{\textrm{ionic}}}{\epsilon_{\textrm{optic}}\epsilon_{\textrm{static}}} ,
    \label{eqn:pekar}
\end{equation}
we can now see that the Fr\"ohlich $\alpha$ is proportional to the ionic dielectric contribution, as would be expected from appreciating that this is the driving force for polaron formation. 

The static dielectric constant is the sum of the high-frequency (`optical') response of the electronic structure, and the lower frequency vibrational response of the ions, $\epsilon_{\textrm{static}}=\epsilon_{\textrm{optic}}+\epsilon_{\textrm{ionic}}$. 
This vibrational contribution is typically calculated\cite{Gonze} by summing the infrared activity of the individual harmonic modes as Lorentz oscillators. 
This infrared activity can be obtained by projecting the Born effective charges along the dynamic matrix (harmonic phonon) eigenvectors.  
The overall dielectric function across the phonon frequency range can be written as 
\begin{equation}
\begin{split}
\epsilon(\Omega) & = \epsilon_{\textrm{optic}} + \sum_{j=1}^{m}\epsilon_{j}(\Omega) \\
& = \epsilon_{\textrm{optic}} + \frac{4\pi}{\Omega_0} \sum_{j=1}^{m} \frac{(U\cdot q)\cdot(U\cdot q)}{\omega_{j}^2-\Omega^2} .
\end{split}
\label{eqn:dielectric}
\end{equation}
Here $U$ are the dynamic matrix eigenvectors, $\Omega$ is the reduced frequency of interest, $\omega_{j}$ is the phonon reduced frequency, $\Omega_0$ is the unit cell volume, $q$ are the Born effective charges, $j$ indexes the $j$th phonon branch and $m$ is the total number of phonon branches.

Considering the isotropic case (and therefore picking up a factor of $\frac{1}{3}$ for the averaged interaction with a dipole), and expressing the static (zero-frequency) dielectric contribution, in terms of the infrared activity of a mode $\epsilon_{j}$ is 
\begin{equation}
\epsilon_{j}(0) 
=  \frac{4\pi}{\Omega_0} \frac{1}{3} \frac{\kappa_j^2 }{\omega_j^2} \, q^2/u
    \label{eqn:epsilon_from_ir}
\end{equation}
where $\kappa$ is the infrared activity in the standard unit of the electron charge ($q$) squared per atomic mass unit ($u$).

This provides a clear route to defining $\alpha_j$ for individual phonon branches, with the simple constitutive relationship that $\alpha=\sum_j \alpha_j$. 
\begin{equation}
\alpha_j = \frac{1}{4\pi\epsilon_0}  \frac{\epsilon_j}{\epsilon_{\textrm{optic}}\epsilon_{\textrm{static}}} \frac{e^2}{\hbar} \left( \frac{m^*}{2\hbar\omega_j} \right)^{\frac{1}{2}}
    \label{eqn:alphai}
\end{equation}


This concept of decomposing $\alpha$ into constituent pieces associated with individual phonon modes is implicit in the effective mode scheme of Hellwarth and Biaggio\cite{Hellwarth1999}, and has also been used by Verdi\cite{carla2017a}, Verbist\cite{verbist1992} and Devreese et al.\cite{devreese2010}.

\subsection{Multiple phonon mode path integral} \label{Sec:multipathintegral}

Verbist and Devreese\cite{verbist1992} proposed an extended Fr\"ohlich model Hamiltonian (Eqn. (\ref{eqn:frohlich})) with a sum over multiple ($m$) phonon branches, 
\begin{equation}
    \begin{aligned}
    \hat{H} &= \frac{p^2}{2m^*} + \sum_{\mathbf{k}, j} \hbar \, \omega_{j} \, a_{\mathbf{k}, j}^\dagger a_{\mathbf{k}, j} \\
    &\quad + \sum_{\mathbf{k}, j} ( V_{\mathbf{k}, j} \, a_{\mathbf{k}, j} \, e^{i\mathbf{k} \cdot \mathbf{r}} + V_{\mathbf{k}, j}^* \, a_{\mathbf{k}, j}^\dagger \, e^{-i\mathbf{k} \cdot \mathbf{r}}) .
    \end{aligned}
\label{eqn:multifrohlich}
\end{equation}

Here the index $j$ indicates the $j$th phonon branch. 
The interaction coefficient is given by,
\begin{equation}
    V_{\mathbf{k}, j} = i \frac{2 \hbar\omega_j}{|\mathbf{k}|} \left(
    \sqrt{\frac{\hbar}{2 m^* \omega_j}} \frac{\alpha_j \pi}{\Omega_0}
    \right)^{\frac{1}{2}} ,
\end{equation}
with $\alpha_j$ as in Eqn. (\ref{eqn:alphai}).

From this Hamiltonian we provide the following extended model action to use within the Feynman variational theory, 
\begin{equation}
    \begin{aligned}
        S_j[\mathbf{r}(\tau)] &=
        \frac{m^*}{2}\int^{\hbar\beta}_0 d\tau \left(\frac{d\mathbf{r}(\tau)}{d\tau}\right)^2 \\
        &\quad -\frac{(\hbar\omega_j)^{\frac{3}{2}}}{2 \sqrt{2 m^*}} \alpha_j \int^{\hbar\beta}_0 d\tau \int^{\hbar\beta}_0 d\sigma \frac{g_{\omega_j}(|\tau - \sigma|)}{|\mathbf{r}(\tau) - \mathbf{r}(\sigma)|} .
    \end{aligned}
\label{eqn:multiaction}
\end{equation}

Here $g_{\omega_j}(\tau)$ is the imaginary-time phonon Green's function for a phonon with frequency $\omega_j$, 
\begin{equation}
    g_{\omega_j}(\tau) = \frac{\cosh{(\omega_j(\tau - \hbar\beta/2))}}{\sinh{(\hbar\omega_j\beta/2)}} .
\end{equation}

This form of action is consistent with Hellwarth and Biaggio's\cite{Hellwarth1999} deduction that inclusion of multiple phonon branches gives the interaction term simply as a sum over terms with phonon frequency $\omega_j$ and coupling constant $\alpha_j$ dependencies.

We now choose a suitable trial action to use with the action in Eqn. (\ref{eqn:multiaction}). We use Feynman's original trial action with two variational parameters, $C$ and $w$, which physically represents a particle (the charge carrier) coupled harmonically to a single fictitious particle (the additional mass of the quasi-particle due to interaction with the phonon field) with a strength $C$ and a frequency $w$.

Clearly the dynamics of this model cannot be more complex than can be arrived at with the original Feynman theory, though the direct variational optimisation (at each temperature) may get closer than using Hellwarth and Biaggio's\cite{Hellwarth1999} effective phonon mode approximation.

\subsection{Multiple phonon mode free energy} \label{Sec:multifreeenergy}

We extend Hellwarth and Biaggio's $A$, $B$ and $C$ equations (Eqs. (62b), (62c) and (62e) in Ref.\cite{Hellwarth1999}) (presented here with explicit units), 
\begin{subequations}
    \begin{align}
        A &= \frac{3}{\hbar\beta\omega_0} \left[ \log\left(\frac{w \sinh(v \hbar\beta\omega_0 / 2)}{v \sinh(w \hbar\beta \omega_0 / 2)}\right) \right. \notag\\
        &\left. \qquad\qquad\quad - \frac{1}{2} \log(2\pi\hbar\beta\omega_0) \right], \\
        B &= \frac{\alpha\omega_0}{\sqrt{\pi}} \int_0^{\frac{\hbar\beta}{2}} d\tau\ g_{\omega_0}(\tau) \left[D(\tau)\right]^{-\frac{1}{2}} \\
        C &= \frac{3}{4} \frac{v^2 - w^2}{v} \left( \coth(\frac{v \hbar\beta\omega_0}{2}) - \frac{2}{v \hbar\beta\omega_0} \right)
    \end{align}
\end{subequations}
to multiple phonon modes, where $D(\tau)$ is given in Eqn. (\ref{eqn:D}). Hellwarth and Biaggio's $B$ is a symmetrised (for ease of computation) version of the equivalent term from \=Osaka\cite{Osaka1959}, although here we have unsymmetrised the integral in $B$ to condense the notation. Compared to \=Osaka, $B$ and $C$ are related to the expectation value of the model action $\langle S \rangle_0$ and trial action $\langle S_0 \rangle_0$, respectively and $A$ is the free energy derived from the trial partition function $F_0 = -\log(Z_0) / \beta$. Following the procedure of \=Osaka\cite{Osaka1959}, from the multiple phonon action in Eqn. (\ref{eqn:multiaction}) we derive the phonon mode dependent $A_j$ and $C_j$ equations,
\begin{subequations}
\begin{align}
    A_j &= \frac{3}{\hbar \omega_j \beta} \left[\log\left(\frac{v \sinh (w \hbar \omega_j \beta / 2)}{w \sinh (v \hbar \omega_j \beta / 2)}\right) \right. \notag\\
    &\left.\qquad\qquad\quad - \frac{1}{2} \log \left(2\pi\hbar \omega_j \beta\right) \right] , \label{eqn:A} \\
    C_j &= \frac{3}{4}\frac{v^2-w^2}{v} \left( \coth \left( \frac{v \hbar \omega_j \beta}{2} \right) - \frac{2}{v \hbar \omega_j \beta} \right) . \label{eqn:C}
\end{align}
\end{subequations}
Similarly, we derive a multiple phonon mode extension to Hellwarth and Biaggio's B expression, 
\begin{equation}
\begin{gathered}
    B_j = \frac{\alpha_j \omega_j}{\sqrt{\pi}} \int_0^{\frac{\hbar\beta}{2}} d\tau g_{\omega_j}(\tau) \left[ D_j(\tau) \right]^{-\frac{1}{2}} ,
\label{eqn:B}
\end{gathered}
\end{equation}
where,
\begin{equation}\label{eqn:multi_D}
\begin{aligned}
    D_j(\tau) &= 2 \frac{v^2 - w^2}{v^3} \frac{\sinh{(v \omega_j \tau/2)\sinh{(v \omega_j[\hbar \beta - \tau]/2)}}}{\sinh(v \hbar \omega_j\beta/2)}\\
    &\quad + \left( 1 - \frac{v^2-w^2}{v^2} \right) \tau \omega_j \left(1 - \frac{\tau}{\hbar\beta}\right).
\end{aligned}
\end{equation}
These are similar to Hellwarth and Biaggio's single mode versions, but with the single effective phonon frequency $\omega_0$ substituted with the branch dependent phonon frequencies $\omega_j$. There are $m$ with index $j$ phonon branches.

Summing $A_j$ in Eqn. (\ref{eqn:A}), $B_j$ in Eqn. (\ref{eqn:B}), and $C_j$ in Eqn. (\ref{eqn:C}), we obtain a generalised variational inequality for the contribution to the free energy of the polaron from the $j$th phonon branch with phonon frequency $\omega_j$ and coupling constant $\alpha_j$, and $2$ variational parameters $v$ and $w$, 

\begin{equation}\label{eqn:multi_feynman_jensen}
        F(\beta) \leq - \sum_{j = 1}^m \hbar \omega_j (A_j + C_j + B_j).
\end{equation}

Here we have taken care to write out the expression explicitly, rather than use `polaron' units. 
The entire sum on the RHS of Eqn. (\ref{eqn:multi_feynman_jensen}) must be minimised simultaneously to ensure we obtain a single pair of $v$ and $w$ parameters that give the lowest upper-bound for the total model free energy $F$.

We obtain variational parameters $v$ and $w$ that minimise the free energy expression and will be used in evaluating the polaron mobility. When we consider only one phonon branch ($m = 1$) this simplifies to Hellwarth and Biaggio's form of \=Osaka's free energy. Feynman's original athermal version can then be obtained by taking the zero-temperature limit ($\beta \rightarrow \infty$).

\subsection{Multiple phonon mode complex mobility} \label{Sec:multimobility}

To generalise the frequency-dependent mobility in Eqn. (\ref{eqn:freq_dep_mobility}), we follow the same procedure as FHIP, but use our generalised polaron action $S$ (Eqn. (\ref{eqn:multiaction})) and trial action $S_0$ (Eqn. (\ref{eqn:trial_action})). The result is a memory function akin to FHIP's $\chi$ (Eqn. (\ref{eqn:fhip_chi})) that now includes multiple ($m$) phonon branches $j$,
\begin{equation}\label{eqn:multichi}
    \begin{gathered}
        \chi_{\textrm{multi}}(\Omega) = \sum_{j=1}^m \frac{\alpha_j \omega_j^{2}}{3\sqrt{\pi}} \int_0^{\infty} dt\ \left[1 - e^{i\Omega t}\right] \textrm{Im} S_j(t)
    \end{gathered} .
\end{equation}
Here, 
\begin{equation}
    S_j(\Omega) = g_{\omega_j}(t) [D_j(t)]^{-\frac{3}{2}} ,
\end{equation}

where $D_j(t)$ is $D_j(\tau = -it)$ from Eqn. (\ref{eqn:multi_D}) rotated back to real-time to give a generalised version of $D(u)$ in Eqn. (35c) in FHIP,
\begin{equation}
    \begin{gathered}
         D_j(t) = 2 \frac{v^2-w^2}{v^3} \frac{\sin(v \omega_j t/2) \sin(v\omega_j[t-i\hbar\beta]/2)}{\sinh(v\omega_j\hbar\beta/2)} \\
        -i \left(1-\frac{v^2-w^2}{v^2}\right) t \omega_j \left(1 - \frac{t}{i\hbar\beta}\right).
    \end{gathered}
\end{equation}

The new multiple-phonon frequency-dependent mobility $\mu_{\textrm{multi}}(\Omega)$ is then obtained from the real and imaginary parts of the generalised $\chi_{\textrm{multi}}(\Omega)$ using Eqn. (\ref{eqn:freq_dep_mobility}).

\section{Comparison between effective-mode and multiple-mode theories} \label{Sec:singlevsmulti}

\begin{table}[!tbp]
\begin{ruledtabular}
\begin{tabular}{lrrrr}
    Material & $\epsilon_{\text{optical}}$ & $\epsilon_{\text{static}}$ & $f$ & $m^*$ \\
    \colrule
    \ce{MAPbI3}-e & 4.5 & 24.1 & 2.25 & 0.12 \\
    \ce{MAPbI3}-h & 4.5 & 24.1 & 2.25 & 0.15 \\
\end{tabular}
\end{ruledtabular}
\caption{
    \label{tab:Params}
    Parameters of the Feynman polaron model (single effective phonon mode) as used in this work.  
    Relative high frequency ($\epsilon_{\text{optical}}$) and static
    ($\epsilon_{\text{static}}$)
    dielectric constants are given in units of the permittivity of free space
    ($\epsilon_0$). Frequency (f) is in \si{\tera\hertz}. Effective mass
    ($m^*$) is in units of the bare electron mass. These data are as in Ref.\cite{Frost2017}.
    }

\end{table}

\begin{table}[!tbp]
   \sisetup{round-mode = figures, round-precision =3}
    \begin{tabular*}{\columnwidth}{S[table-format=3.2]@{\extracolsep{\fill}}  S[table-format=3.2] S[table-format=3.2] S[table-format=3.2]} 
   \toprule   \vspace{-6pt}\\
        {Base frequency} & {Polaron frequency} & {i.r. activity} & {$\alpha_j$} \\
    \colrule
4.016471586720514  &  10.760864419751513  &  0.08168931020200264  &  0.034  \\
3.887605410774121  &  10.415608286921941  &  0.006311654262282101  & 0.003  \\
3.5313112232401513  &  9.461030777081  &  0.05353548710183397  &  0.031  \\
2.755392921480459  &  7.382203262491912  &  0.021303020776321225  &  0.023  \\
2.4380741812443247  &  6.532048128115507  &  0.23162784335484837  &  0.336  \\
2.2490917637719408  &  6.0257295526621535  &  0.2622203718355982  &  0.465  \\
2.079632190634424  &  5.571716259703516  &  0.23382298607799906  &  0.505  \\
2.0336707697261187  &  5.4485771789818624  &  0.0623239656843172  &  0.142  \\
1.5673011873879714  &  4.199087487176367  &  0.0367465760261409  &  0.161  \\
1.0188379384951798  &  2.7296537981481457  &  0.0126328938653956  &  0.162  \\
1.0022960504442775  &  2.6853350445558144  &  0.006817361620021601  &  0.091  \\
0.9970130778462072  &  2.6711809915185523  &  0.0103757951973341  &  0.141  \\
0.9201781906386209  &  2.4653262291740656  &  0.01095811116040592  &  0.182  \\
0.800604081794174  &  2.144965249242841  &  0.0016830270365341532  &  0.040  \\
0.5738689505255512  &  1.537500225752314  &  0.00646428491253749  &  0.349  \\
\botrule
\end{tabular*}

\caption{\label{tab:simulatedspectra}Infrared activity of phonon modes in
\ce{MAPbI$_3$} taken from Ref. \onlinecite{Brivio2015}, scaled to their ground-state polaron value by the multimodal $w=2.6792$ factor for \ce{MAPbI3}-e of this work (Table \ref{tab:Results}). 
} 
\end{table}

\begin{table}[!tbp]
    \sisetup{round-mode = figures, round-precision =3}
    \begin{tabular*}{\columnwidth}{l@{\extracolsep{\fill}}cccS}
\toprule   \vspace{-6pt}\\
    Material & $\alpha$ & $v$ & $w$ & {$E_b$} \\ 
\colrule
        \ce{MAPbI3}-e & 2.39  & 3.3086 & 2.6634 & \SI{-23.041730}{\meV} \\
        \ce{MAPbI3}-h & 2.68  & 3.3586 & 2.6165 & \SI{-25.879823}{\meV} \\
\colrule
        \ce{MAPbI3}-e & 2.66  & 3.2923 & 2.6792 &  \SI{-19.516889}{\meV} \\
        \ce{MAPbI3}-h & 2.98  & 3.3388 & 2.6349 &  \SI{-21.915437}{\meV} \\
\botrule
\end{tabular*}
\caption{\label{tab:Results} Athermal 0 K results. Dielectric electron-phonon coupling ($\alpha$), Feynman athermal variational parameters ($v$ and $w$) and polaron binding energy ($E_b$) for an effective phonon mode (top rows) and for multiple explicit phonon modes (bottom rows).   
}
\end{table}

\begin{table}[!tbp]
    \sisetup{round-mode = figures, round-precision =3}
    \begin{tabular*}{\columnwidth}{l@{\extracolsep{\fill}}cccccccc}
\toprule   \vspace{-6pt}\\
   {Material} & {$\alpha$} & {$v$} & {$w$} & {$F$} & {$\mu$} & {$M$} & {$r_f$} \\ 
\colrule
        \ce{MAPbI3}-e & 2.39 & 19.9 & 17.0 & -35.5 & 136 & 0.37 & 43.6 \\
        \ce{MAPbI3}-h & 2.68 & 20.1 & 16.8 & -43.6 & 94 & 0.43 & 36.9 \\
\colrule
        \ce{MAPbI3}-e & 2.66 & 35.2 & 32.5 & -42.8 & 160 & 0.18 & 44.1 \\
        \ce{MAPbI3}-h & 2.98 & 35.3 & 32.2 & -50.4 & 112 & 0.20 & 37.2 \\
\botrule
\end{tabular*}
\caption{\label{tab:Results300K} 300 K Results. Dielectric electron-phonon coupling ($\alpha$), Feynman thermal variational parameters ($v$ and $w$), polaron free energy ($F$, meV), dc mobility ($\mu$, cm$^2$V$^{-1}$s$^{-1}$), polaron effective mass ($M$, $m^*$) and Schultz polaron radius ($r_f$, \r{A}) for an effective phonon mode (top rows) and for multiple explicit phonon modes from Table \ref{tab:simulatedspectra} (bottom rows).   
}
\end{table}

Having extended the Feynman theory with explicit phonon modes in the \emph{model} action, we must now try and answer what improvement this makes. 

Halide perovskites are relatively new semiconductors of considerable technical interest. 
They host strongly interacting large polarons due to their unusual mix of light effective mass yet strong dielectric electron-phonon coupling.
Recently the coherent charge-carrier dynamics upon photo-excitation are being measured, the Terahertz spectroscopy showing rich transient vibrational features\cite{Guzelturk2018}.

Therefore, we choose to use this system as representative of the more complex systems which could be modelled with our extended theory. 

In what follows, we take the materials data from our 2017 paper\cite{Frost2017}, which we reproduce here in Table \ref{tab:Params}.

\subsection{Free energy} \label{Sec:compfreeenergy}

\begin{figure}[!tbp]
\centering
    \includegraphics[width=\columnwidth]{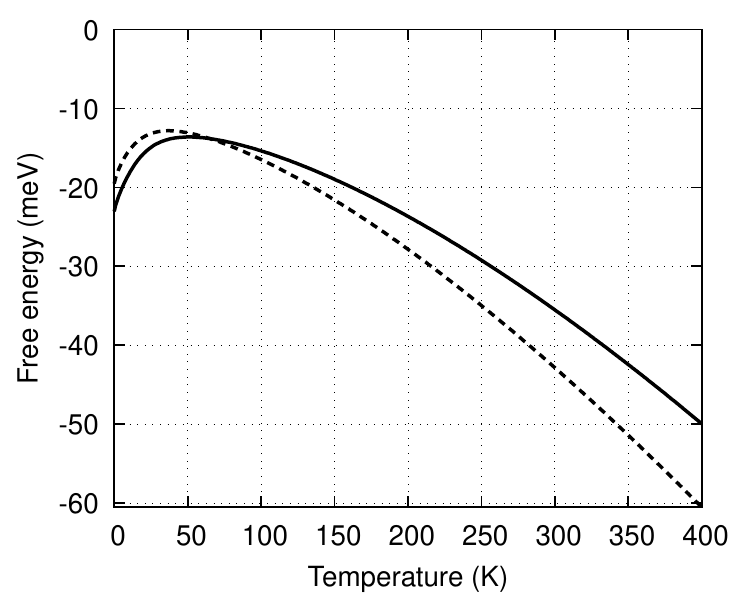}
    \caption{\label{fig:energycomparison} Comparison of the polaron free energy as a function of temperature for MAPbI$_3$ with the single effective phonon mode approach (solid) and the explicit multiple phonon mode approach (dashed).
}
\end{figure}

\begin{figure}[!tbp]
\centering
    \includegraphics[width=\columnwidth]{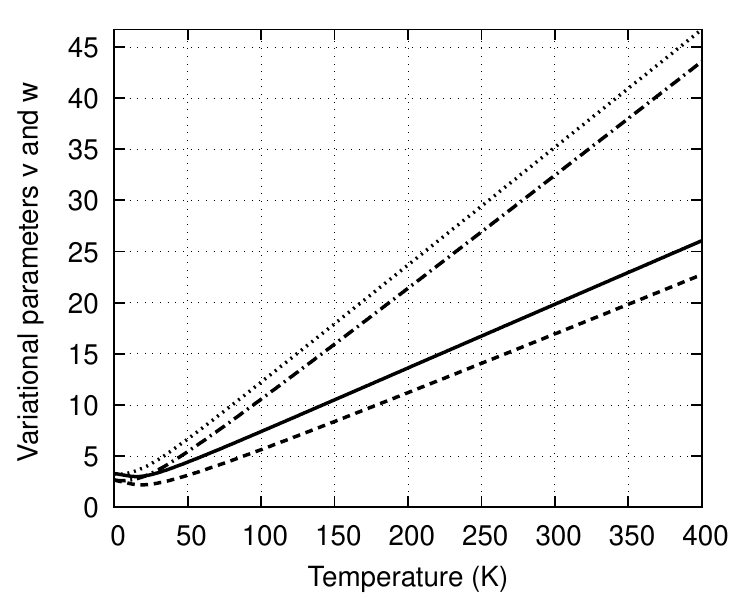}
    \caption{\label{fig:varparamscomparison} Comparison of the two polaron variational parameters ($v$ and $w$) for MAPbI$_3$ in the single effective phonon mode approach ($v$, solid; $w$ dashed) and the explicit multiple phonon mode approach ($v$, dots; $w$ dot-dashes).
}
\end{figure}

We compare the polaron free energy and variational parameters evaluated by our explicit phonon frequency method presented in Eqn. (\ref{eqn:multi_feynman_jensen}) to Hellwarth and Biaggio's effective phonon frequency scheme (scheme `B' in Eqs. (58) and (59) in Ref. \onlinecite{Hellwarth1999}),
\begin{subequations}
    \begin{align}
        \frac{\kappa_{\text{eff}}^2}{\omega_{\text{eff}}^2} &= \sum_{j=1}^m \frac{\kappa_j^2}{\omega_k^2} \label{eqn:b1}\\
        \kappa_{\text{eff}}^2 &= \sum_{j=1}^m \kappa_j^2, \label{eqn:b2}
    \end{align}
\end{subequations}
that use an effective LO phonon mode frequency $\omega_{\text{eff}}$ and associated  infrared oscillator strength $\kappa_{\text{eff}}$ derived from sums over the phonon modes $j$. 
We apply both of these methods to the 15 solid-state optical phonon branches of MAPbI$_3$, of which the frequencies and infrared activities are shown in Table \ref{tab:simulatedspectra}.

Using the Hellwarth and Biaggio\cite{Hellwarth1999} effective phonon frequency `B' scheme, the effective phonon frequency for \ce{MAPbI3} is $\omega_0 = 2.25 \cdot 2\pi$ THz and the Fr\"ohlich alpha for \ce{MAPbI3}-e is $\alpha = 2.39$ and \ce{MAPbI3}-h is $\alpha = 2.68$, as in our previous work\cite{Frost2017} (values from bulk dielectric constants).

Using Eqn. (\ref{eqn:alphai}), we calculated the partial Fr\"ohlich alpha $\alpha_j$ parameters for each of the 15 phonon branches in \ce{MAPbI3}, which are given in Table \ref{tab:simulatedspectra}. For \ce{MAPbI3}-e the partial Fr\"ohlich alphas sum to $\alpha = 2.66$ and for \ce{MAPbI3}-h they sum to $\alpha = 2.98$. These 15 partial alphas $\alpha_j$ and corresponding phonon frequencies $\omega_j$ were then used in the variational principle for the multiple phonon dependent free energy in Eqn. (\ref{eqn:multi_feynman_jensen}). From Eqn. (\ref{eqn:multi_feynman_jensen}), we variationally evaluate a $v$ and $w$ parameter.

Fig. \ref{fig:energycomparison} shows the polaron free energy comparison. The explicit multiple phonon mode approach predicts a higher free energy at temperatures $T < 65$K and a lower free energy at temperatures $T > 65$K. See Table \ref{tab:Results} for our athermal results, where we find new multiple-mode estimates for the polaron binding energy $E_b$ (at \SI{0}{\kelvin}) for \ce{MAPbI3}-e as $E_b = -19.52$ meV and \ce{MAPbI3}-h as $E_b = -21.92$ meV. Also see Table \ref{tab:Results300K} for our thermal results at $T = 300$ K, where where we find new multiple-mode estimates for the polaron free energy $F$ for \ce{MAPbI3}-e at $300$ K as $F = -42.84$ meV and \ce{MAPbI3}-h as $F = -50.40$ meV. These are to be compared to our previous results in Ref. \onlinecite{Frost2017}, which are also provided in Tables (\ref{tab:Results}) and (\ref{tab:Results300K}).

Fig. \ref{fig:varparamscomparison} shows the comparison in polaron variational parameters $v$ and $w$. That we have different trends for the polaron free energy and variational $v$ and $w$ parameters, shows that we find quite a different quasi-particle solution from our multiple phonon scheme compared to the single effective frequency scheme.

\subsection{DC mobility} \label{Sec:compmobility}

\begin{figure}[!tbp]
\centering
    \includegraphics[width=\columnwidth]{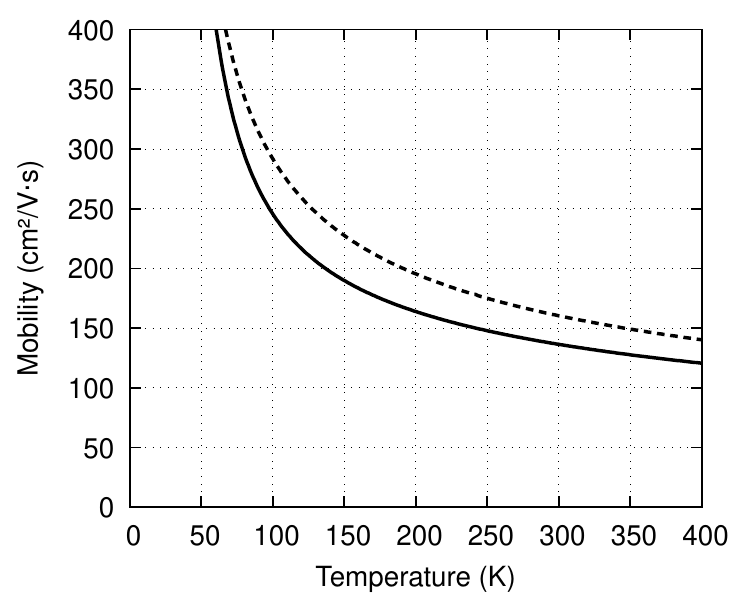}
    \caption{\label{fig:mobilitycomparison} 
    Comparison of the temperature-dependent mobility predicted for \ce{MAPbI3} by the single effective phonon mode approach (solid) and the explicit multiple phonon mode approach (dashed).
}
\end{figure}

\begin{figure}[!tbp]
\centering
    \includegraphics[width=\columnwidth]{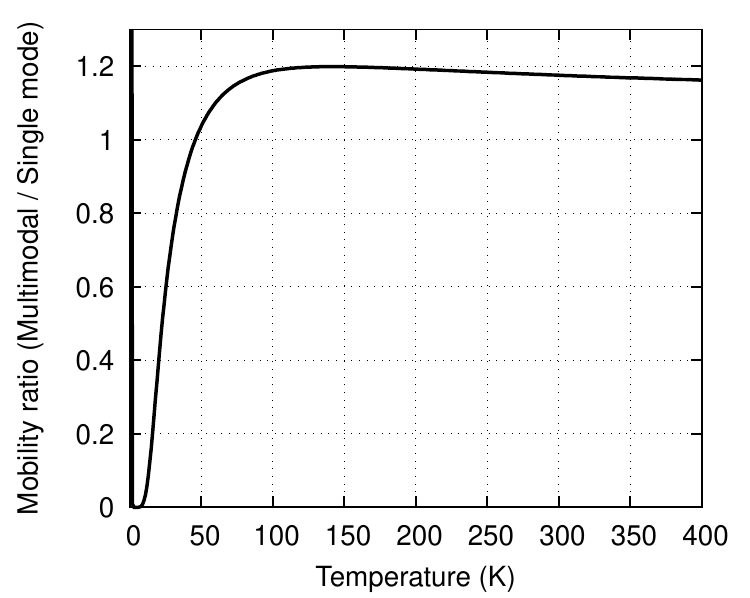}
    \caption{\label{fig:mobilityratiocomparison} 
    Ratio of the temperature-dependent mobility predicted for \ce{MAPbI3} by the explicit multiple and single effective phonon mode approachs (solid). The multiple mode approach gives up to $20$ \% correction, maximised at $T = 140$ K. 
}
\end{figure}

We calculate the zero-frequency (direct current, dc) electron-polaron mobility $\mu$ in MAPbI$_3$ using the effective phonon mode and explicit multiple phonon mode approaches. Both approaches have the same relationship between the mobility and the memory function (Eqns. (\ref{eqn:freq_dep_mobility}) and (\ref{eqn:dcmobility})), but the effective mode approach uses the memory function $\chi(\Omega)$ from Eqn. (\ref{eqn:fhip_chi}) (the FHIP\cite{Feynman1962} memory function, Eqn. (35) ibid.), whereas the multiple phonon mode approach uses our $\chi_{\textrm{multi}}(\Omega)$ from Eqn. (\ref{eqn:multichi}) (with a sum over the phonon modes). Fig. \ref{fig:mobilitycomparison} shows temperatures \SIrange{0}{400}{\kelvin}. In Fig. \ref{fig:mobilityratiocomparison} we see that the multiple mode approach corrects the single effective mode approach by up to 20\%, with this correction maximised at T$ = $\SI{140}{\kelvin}. The multiple mode mobility slowly approaches the single mode mobility towards higher temperatures. We assume the divergence towards zero temperature to be numerical error due to the ratio of large floating-point numbers as both mobility values diverge to positive infinity.

\subsection{Complex conductivity and impedance} \label{Sec:compconduct}

\begin{figure}[!tbp]
\centering
    \includegraphics[width=\columnwidth]{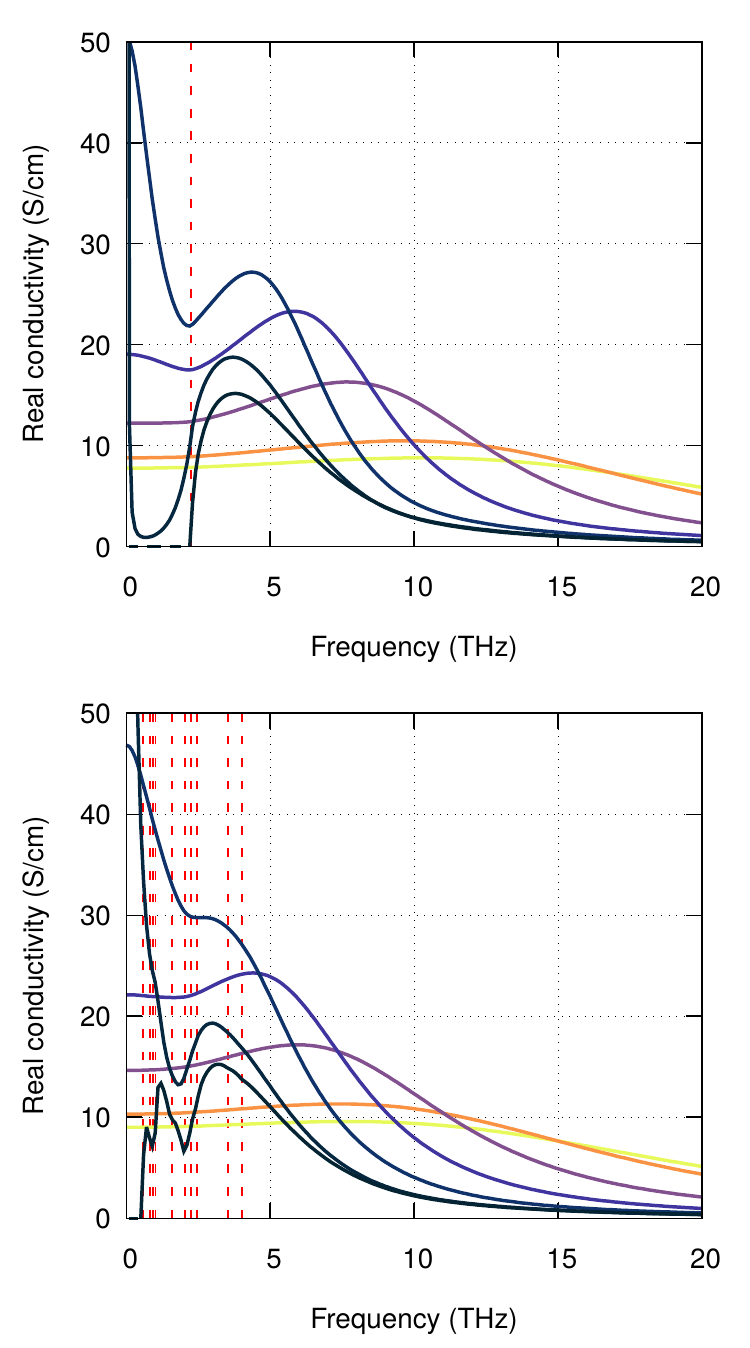}
    \caption{\label{fig:realconductivitycomparison} Real component of the complex conductivity for MAPbI$_3$ for temperatures $T = 0$K, $10$K, $40$K, $80$K, $150$K, $300$K \& $400$K starting with the black curve and finishing with the yellow curve. (Top) Single effective phonon mode prediction. (Bottom) Explicit multiple phonon mode prediction. The red vertical dashed lines indicate the frequencies of the phonon modes.
}
\end{figure}

\begin{figure}[!tbp]
\centering
    \includegraphics[width=\columnwidth]{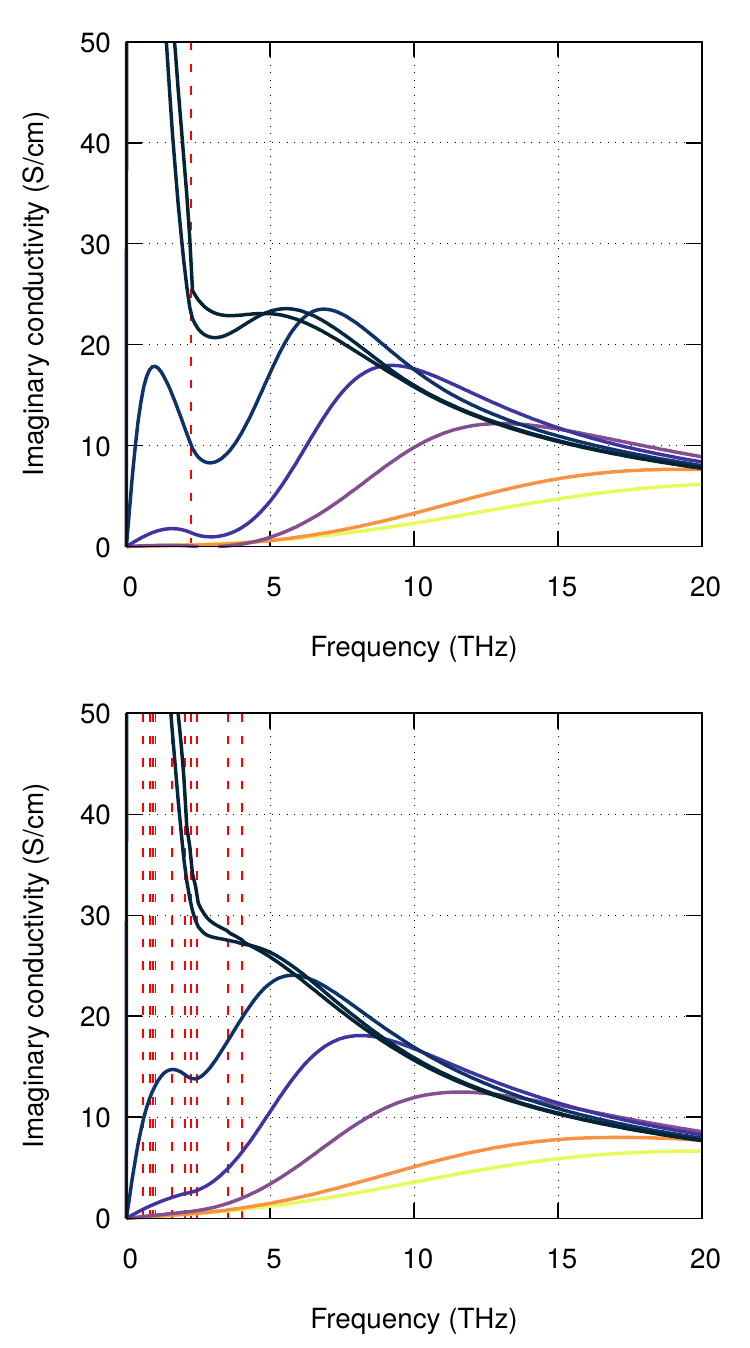}
    \caption{\label{fig:imagconductivitycomparison} Imaginary component of the complex conductivity for MAPbI$_3$ for temperatures $T = 0$K, $10$K, $40$K, $80$K, $150$K, $300$K \& $400$K starting with the black curve and finishing with the yellow curve. (Top) Single effective phonon mode prediction. (Bottom) Explicit multiple phonon mode prediction. The red vertical dashed lines indicate the frequencies of the phonon modes.
}
\end{figure}

\begin{figure}[!tbp]
\centering
    \includegraphics[width=\columnwidth]{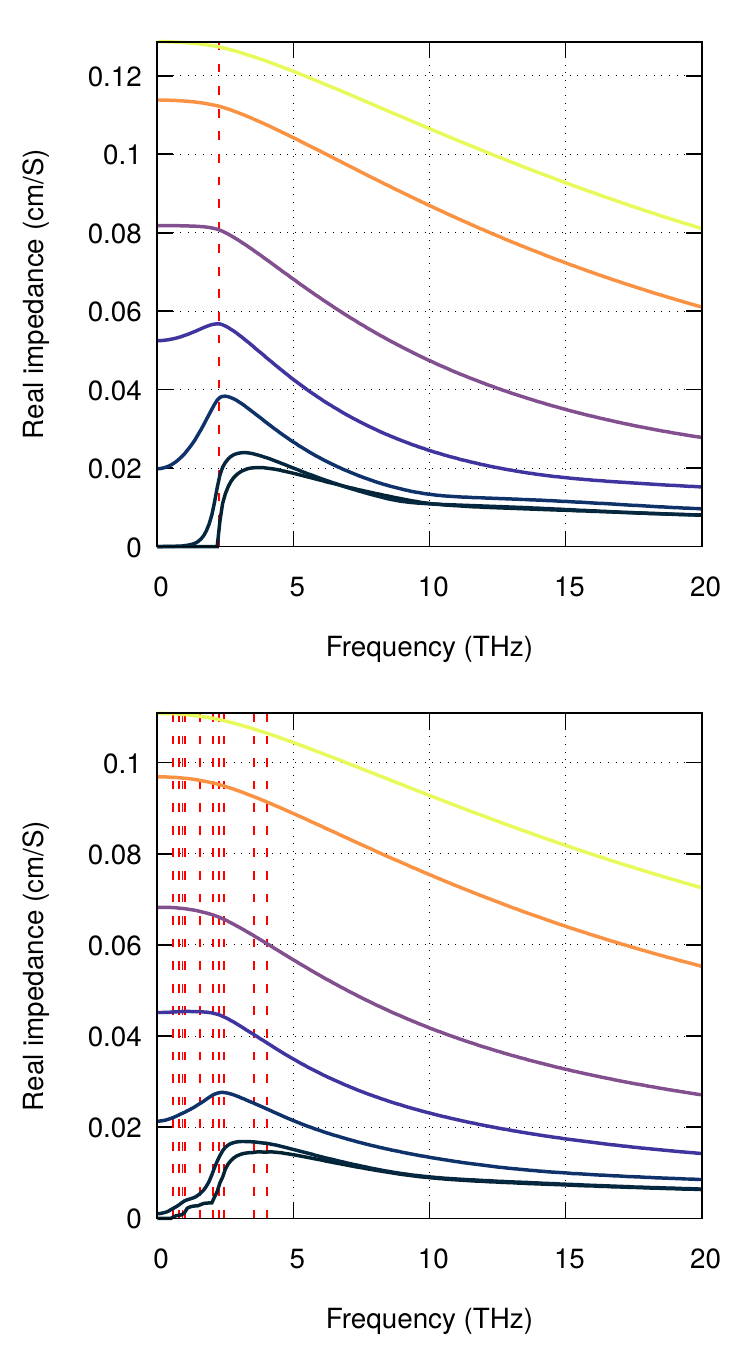}
    \caption{\label{fig:realimpedancecomparison} Real component of the complex impedance for MAPbI$_3$ for temperatures $T = 0$K, $10$K, $40$K, $80$K, $150$K, $300$K \& $400$K starting with the black curve and finishing with the yellow curve. (Top) Single effective phonon mode prediction. (Bottom) Explicit multiple phonon mode prediction. The red vertical dashed lines indicate the frequencies of the phonon modes.
}
\end{figure}

\begin{figure}[!tbp]
\centering
    \includegraphics[width=\columnwidth]{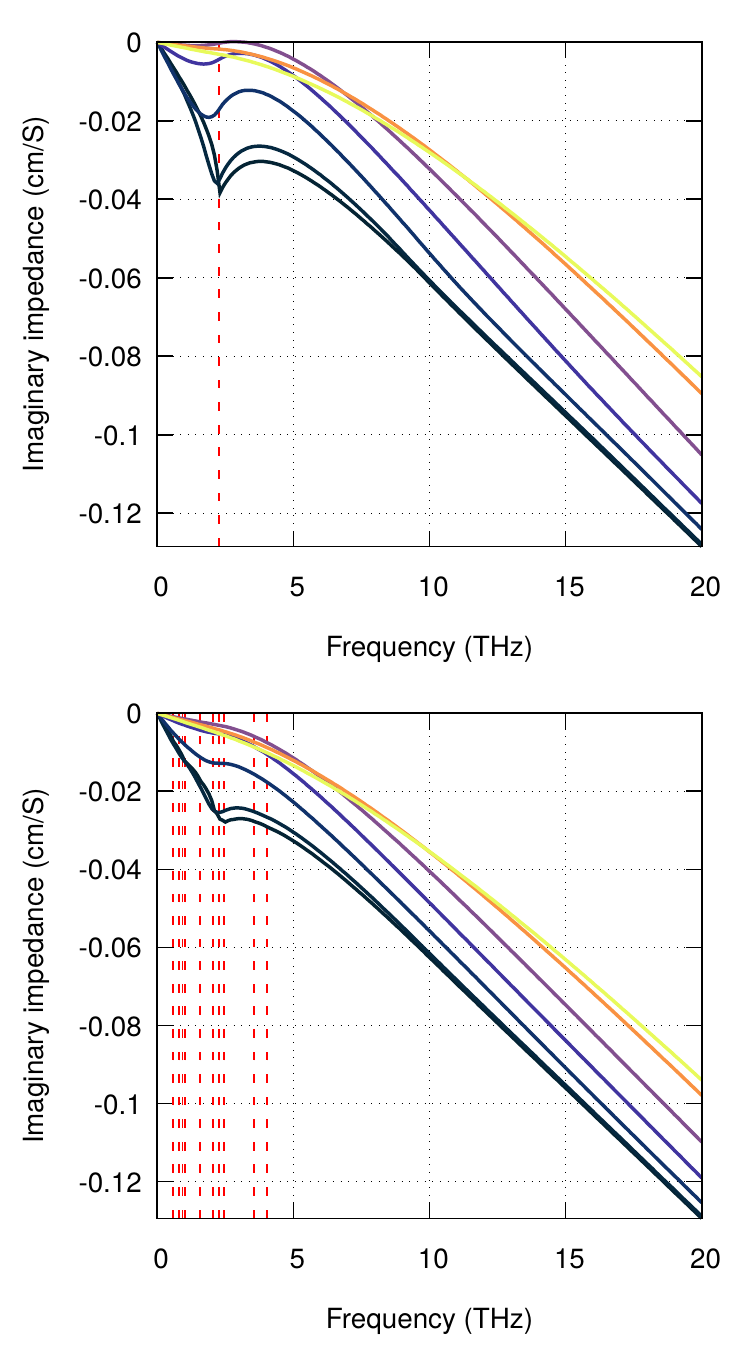}
    \caption{\label{fig:imagimpedancecomparison} Imaginary component of the complex impedance for MAPbI$_3$ for temperatures $T = 0$K, $10$K, $40$K, $80$K, $150$K, $300$K \& $400$K starting with the black curve and finishing with the yellow curve. (Top) Single effective phonon mode prediction. (Bottom) Explicit multiple phonon mode prediction. The red vertical dashed lines indicate the frequencies of the phonon modes.
}
\end{figure}

\begin{figure}[!tbp]
\centering
    \includegraphics[width=\columnwidth]{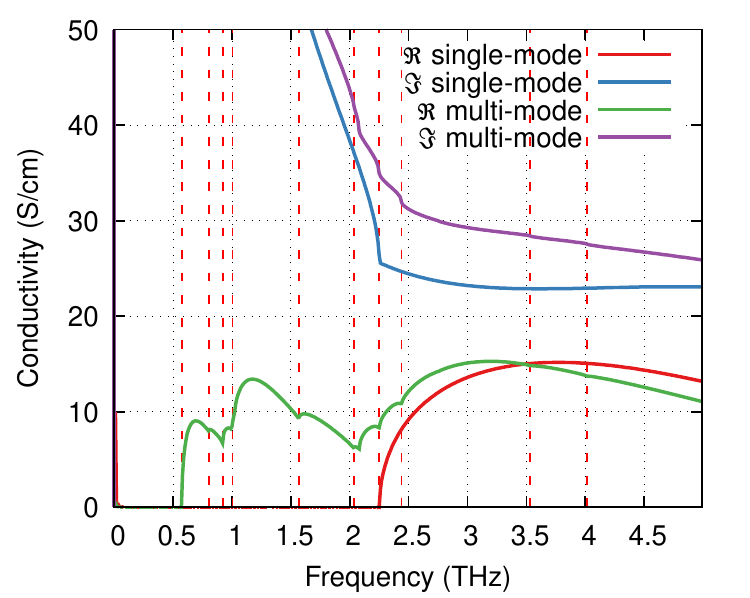}
    \caption{\label{fig:zeroconductivitycomparison} Comparison between the real and imaginary components of the complex conductivity predicted for MAPbI$_3$ by the single effective phonon mode approach and the explicit multiple phonon mode approach.
}
\end{figure}

We calculate the complex impedance $z_{\textrm{multi}}(\Omega)$ for the polaron in MAPbI$_3$ using Eq. (\ref{eqn:compleximpedance}), where the only difference between the effective mode and multiple mode approaches is in the form of the memory function $\chi_{\textrm{multi}}(\Omega)$ as described for the polaron mobility above. The complex conductivity $\sigma_{\textrm{multi}}(\Omega)$ is the reciprocal of the complex impedance, $\sigma_{\textrm{multi}} (\Omega) = 1 / z_{\textrm{multi}}(\Omega)$.

We show in Fig. \ref{fig:realconductivitycomparison} the real component, and in Fig. \ref{fig:imagconductivitycomparison} the imaginary component, of the complex conductivity for the single effective mode approach (top) and the explicit multiple mode approach (bottom) for temperatures $T = 0$K, $10$K, $40$K, $80$K, $150$K, $300$K and $400$K (starting with the black solid line through to the yellow solid line) and for frequencies $0 \leq \Omega \leq 20$ THz. The vertical dashed red lines show the LO phonon modes of MAPbI$_3$. The difference between the two approaches is largest at low temperatures $T = 0$K and $10$K where the multiple phonon approach has more structure due to the extra phonon modes. At higher temperatures, the structure attenuates and the two approaches show similar frequency dependence of the complex conductivity at $T = 300$K and $400$K. These features are further reflected in the real and imaginary components of the complex impedance as shown in Fig. \ref{fig:realimpedancecomparison} and Fig. \ref{fig:imagimpedancecomparison} respectively.

In Fig. \ref{fig:zeroconductivitycomparison} we  specifically show the real and imaginary components of the complex conductivity at zero temperature $T = 0$K over frequencies $0 \leq \Omega \leq 5.0$ THz for both approaches. Again, the vertical dashed red lines show the longitudinal optical (LO) phonon modes of MAPbI$_3$ used in the calculation and are shown in Table \ref{tab:simulatedspectra}. The single effective mode conductivity shows a peak in the real component at frequencies above the effective mode frequency$\Omega \geq 2.25$ THz. 
Whereas, the real component of the multiple mode conductivity shows peaks at frequencies at and above the LO phonon mode frequencies in MAPbI$_3$. 
The imaginary components of both approaches show some structure changes at their respective LO phonon mode frequencies, but are harder to discern at zero temperature. 
The most prominent modes in MAPbI$_3$ appear at the large electron-phonon coupled modes $\omega_0 = 0.58$, $1.00$ and $2.44$ THz.

\section{Simulated polaron vibrational mode spectra}\label{section:spectra}

\begin{figure}[!tbp]
\centering
    \includegraphics[width=\columnwidth]{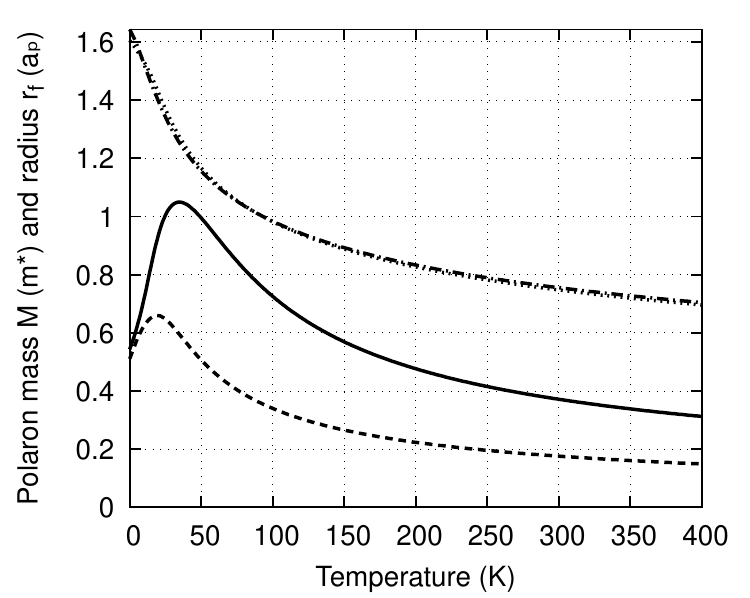}
    \caption{\label{fig:massradiuscomparison} Comparison of polaron effective mass $M$ (in units of effective band mass $m^*$) and Schultz polaron radius $r_f$ (in units of characteristic polaron length $a_p$) for MAPbI$_3$ in the single effective phonon mode approach ($M$, solid; $r_f$ dashed) and the explicit multiple phonon mode approach ($M$, dots; $r_f$, dot-dashes).
}
\end{figure}

\begin{figure}[!tbp]
    \includegraphics[width=\columnwidth]{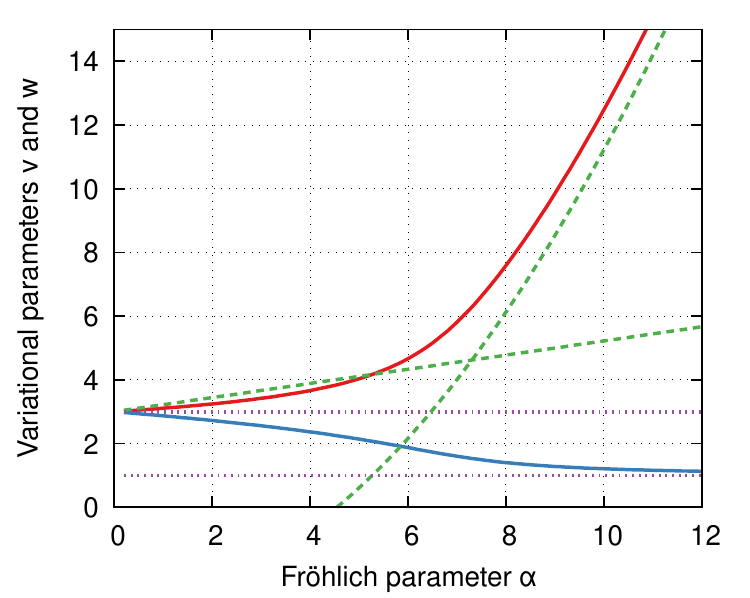}
    \caption{\label{fig:athermal_vs_asymptotic}Numeric Feynman variational solution with the original athermal actions. 
    Blue circles are the value for $w$, red crosses the value for $v$
    Also shown are the asymptotic approximations, as presented in the original paper\cite{Feynman1955} and summarised (often with typos) in textbooks\cite{FeynmanStatMech1972}. 
    The strong ($v=\frac{4\alpha^2}{9\pi} - \frac{3}{2} (2 \mathrm{log}(2) + c)
    - \frac{3}{4}$) and weak ($v=3(1+2\alpha(1-P(w))/3w)$) coupling
    approximations for $v$ are green lines, where $C \approx 0.5772$ is the
    Euler Mascheroni constant and $P(w) = 2[(w-1)^{\frac{1}{2}} - 1]/w \approx 0.2761$ for $w=3$. The
    weak ($w=3$) and strong ($w=1$) approximations are purple lines.     
}
\end{figure}

\begin{figure}[!tbp]
\centering
 \includegraphics[width=3.68in]{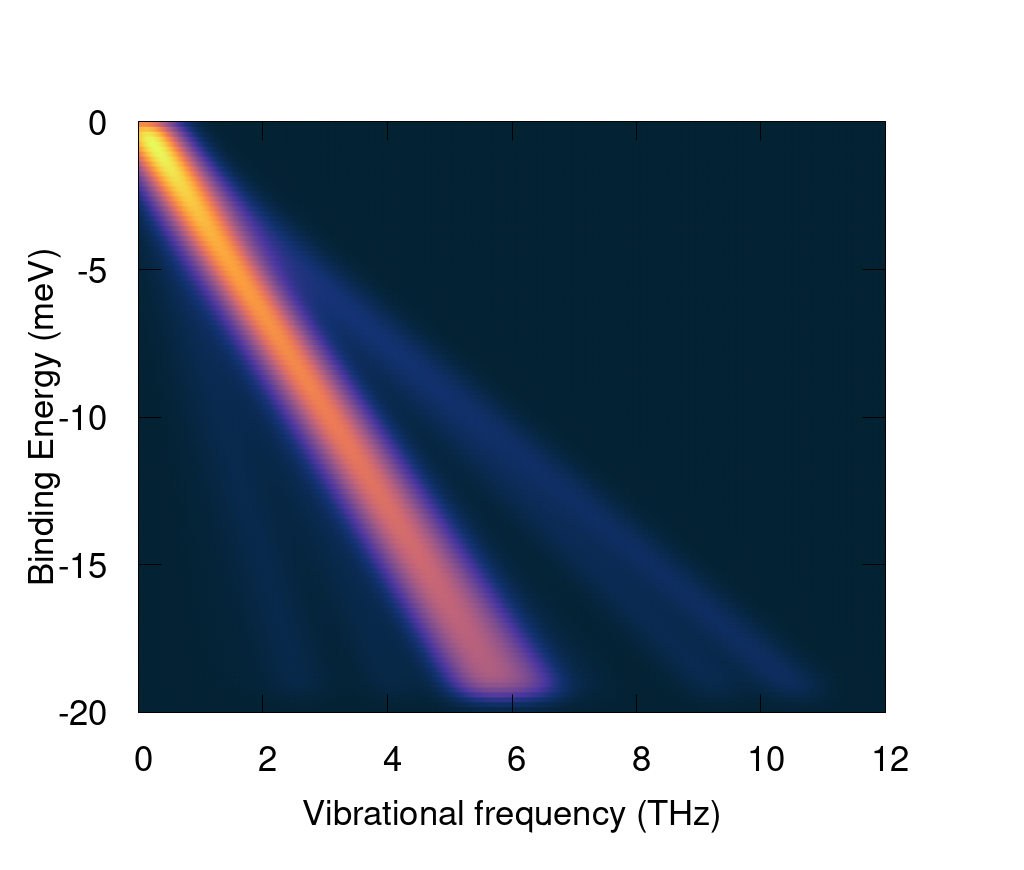}
 \includegraphics[width=3.68in]{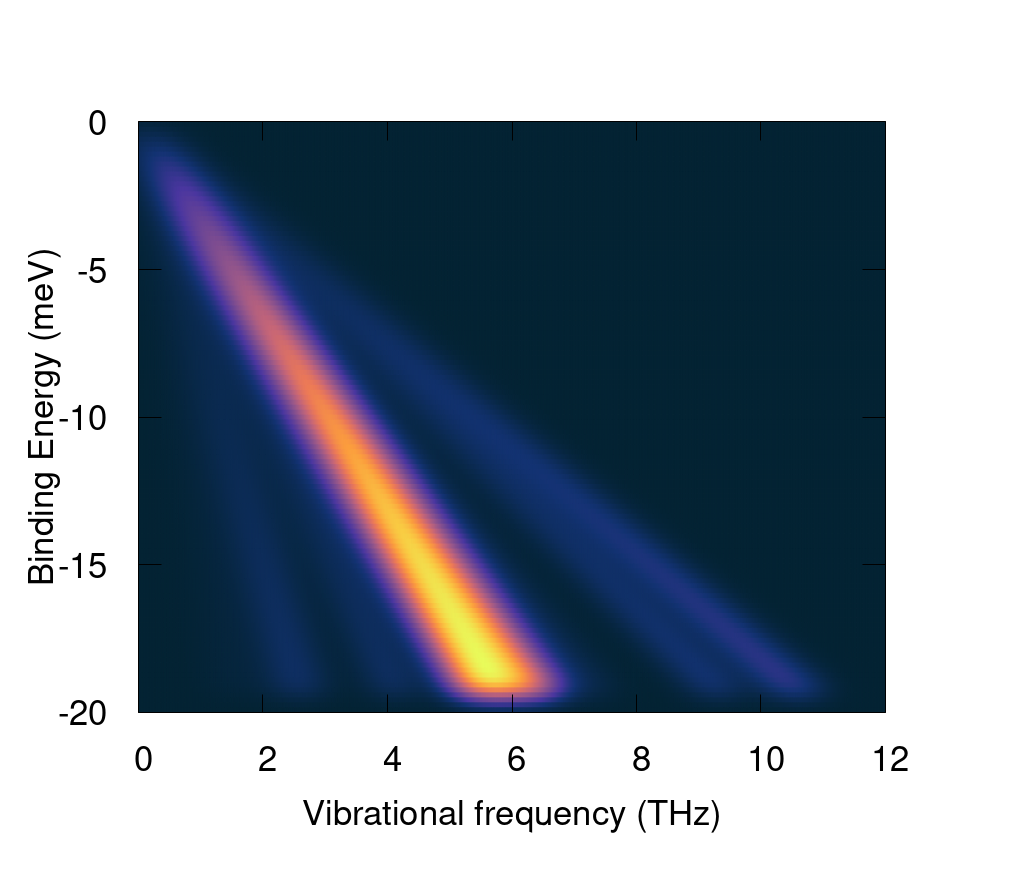}

 \caption{\label{fig:JudgmentDay}Simulated \ce{MAPbI3} polaron vibrational
 spectrum. 
 Data consist of the polaron renormalised vibrational modes (Table
 \ref{tab:simulatedspectra}), with frequency linear varied between 0 at the
 excited band-edge state, to the fully renormalised frequency in the polaron
 ground state (predicted to lie \SI{19.517}{\meV} below the band-edge). 
 These straight lines are weighted by the infrared activity (from Ref. \onlinecite{Brivio2015}, calculated by projecting Born-effective-charges along
 the gamma-point vibrational modes.  These data are then smoothed with a two
 dimensional Kernel density
 estimator, with Gaussian widths of \SI{0.5}{\tera\hertz} (horizontal) and
 \SI{0.5}{\meV} (vertical), to provide a guide to how a noise-less
 low-temperature measurement is predicted to look with this theory. 
 (Top) constant infrared activity assumed across binding energies. 
 (Bottom) infrared activity assumed to attenuate to zero as higher lying
 polaron excited states are accessed. 
 }
\end{figure}

The Feynman polaron quasi-particle has a direct mechanistic interpretation. 
The Lagrangian consists of an effective-mass electron, an additional fictitious particle (mass $M$, in units of the electron effective-mass), coupled by a harmonic restoring force ($k$). This Lagrangian is given by
\begin{equation}\label{eqn:triallagrangian}
    \begin{aligned}
        L = \frac{m^*}{2} \vb{\dot{r}}(t)^2 + \frac{M}{2} \vb{\dot{R}}(t)^2 - \frac{k}{2} \left(\vb{r}(t) - \vb{R}(t)\right)^2.
    \end{aligned}
\end{equation}
The rate of oscillation of this mode is simply $w=\sqrt{\frac{k}{M}}$, expressed as a pre-factor to the material phonon frequency. This oscillation describes the coherent exchange of energy between the electron and the phonon-field. The phonon frequencies are blue-shifted by the electron-phonon coupling. In terms of the variational parameters $v$ and $w$, the spring constant is $k = v^2 - w^2$ and the fictitious mass is $M = (v^2 - w^2) / w^2$.

Following Schultz\cite{Schultz1959}, the size of the polaron is estimated by
calculating the root mean square distance between the electron and the
fictitious particle, given as $r_f = \left(\langle\vb{r}
- \vb{R}\rangle\right)^{\frac{1}{2}} = \sqrt{3v/(v^2 - w^2)}\ a_p$, where the polaron radius is in units of characteristic polaron length $a_p = \sqrt{\hbar/(2m^*\omega_0)}$.

Fig. \ref{fig:massradiuscomparison} shows the comparison in the polaron effective mass $M$ (units of effective electron mass $m^*$) and polaron radius $r_f$ (units of characteristic polaron length $a_p$) applied to \ce{MAPbI3}. At $300$ K we find a new estimates of $M = 0.18$ $m^*$ and $r_f = 0.755$ $a_p$ $= 44.08$ \r{A}, to be compared to $M = 0.37$ $m^*$ and $r_f = 0.747$ $a_p$ $= 43.62$ \r{A} (Table \ref{tab:Results300K}). Whilst the introduction of multiple phonon modes barely alters the polaron size, we note that it practically halves the polaron effective mass.

The original work of Feynman\cite{Feynman1955} provides several asymptotic estimates of this $w$ parameter. The standard approximations, often reproduced in textbooks, are $w=3$ for small $\alpha$ coupling, and $w=1$ for large $\alpha$ coupling. 
Precise work requires a numeric solution, but these limits inform us that the internal polaron mode, as a function of electron-phonon coupling, starts as a harmonic of $3\,\omega_0$ and continuously red-shifts to the phonon fundamental frequency $\omega_0$. 
These parameters as a function of $\alpha$ are shown in Fig. \ref{fig:athermal_vs_asymptotic}.
As there does not seem to be a reference in the literature for numeric values of $w$ (and $v$) as a function of $\alpha$, we provide this in the supplemental information.
From these values, and Eqn. \ref{eqn:alpha}, the predicted polaron vibrational renormalisation can be calculated for an arbitrary material without recourse to further numeric calculation. 

The finite temperature \=Osaka\cite{Osaka1959} action is also described by the Lagrangian which describes the free energy of the polaron state and has the simple mechanistic interpretation of the electron at position $\vb{r}(t)$ coupled by a spring with force constant $k$ to a fictitious particle of mass $M$ at position $\vb{R}(t)$. Practically, the finite temperature action gives rise to a set of $v$ and $w$ parameters which scale almost linearly in temperature (Fig. \ref{fig:varparamscomparison} or see Fig. 3 in Ref. \onlinecite{Frost2017}). 
Na\"ive interpretation of those values as simple harmonic oscillators would suggest infeasible high-frequency oscillations at room temperature, with a strong (almost linear) temperature dependence. It may be possible to disentangle the entropic contribution in this Lagrangian, so calculate the correct temperature dependence of the polaron vibration. 



Each of the individual dielectrically coupled phonon modes will be scaled by this factor. The electron-phonon coupling in the Fr\"ohlich model is linear, and proportional to the infrared activity of the phonon mode. We can therefore simply plot the expected phonon vibrational spectrum from this model, multiplying the phonon frequencies by the scaling factor $w$, and directly
taking the intensity from the infrared activity. 

This rate of vibration is for the ground state of the polaron. The polaron binding energies, indicating where this polaron state is relative to the band edges is given in Table \ref{tab:Results}. 
The states between here and the band edges are a continuum from the fully bound state (where $k$ is some factor of v and w) to a fully unbound state (where $k=0$). We can expect $k$ to linearly decrease as a function of polaron excitation, and so the observed polaron vibrational modes will decrease linearly from these ground state values, to zero at the unbound (band edge) state.

As an example to guide experiment interpretation, we simulate a polaron
vibrational measurement. We choose the archetype methylammonium-lead-halide perovskite material. 
The gamma-point phonon frequencies and infrared activities we take from
a previous set of density-functional-theory lattice-dynamic
calculations\cite{Brivio2015}. 

We plot these modes as a function of energy below the band edge. 
The spring-coupling constant is varied linearly between zero at the band edge,
to the full ground state value (w=2.68). 
This factor scales the vibrational mode. 

The resulting finite set of modes and infrared activities are smoothed with
a two dimensional Kernel density estimator, with a Gaussian width of
\SI{0.5}{\tera\hertz} and \SI{0.5}{\meV}. 
This is intended as a simulation of spectra resolved at low-temperature. 

\section{Discussion} \label{Sec:discussion}

We have shown that the 60 year old FHIP\cite{Feynman1962} mobility theory reproduces much of the `beyond quasiparticle' behaviour exhibited in the recent diagrammatic Monte-Carlo calculations\cite{Mishchenko2019}, including violation of the ``thermal'' Mott-Ioffe-Regel criterion (or Planckian bound\cite{Hartnoll2021}) and, non-monotonic temperature dependence. 

Additionally, we have extended the Feynman variational approach to the polaron problem to include multiple phonon modes in the effective model action. 
Compared to Hellwarth and Biaggio's\cite{Hellwarth1999} effective mode method, we see additional structure in the frequency-dependent mobility, which has recently become something that can be directly measured\cite{Zheng2021} in the Terahertz regime.

\subsection{Violation of the Mott-Ioffe-Regel criterion versus Planckian bound}

The usual MIR criterion puts bounds on transport coefficients of the Boltzmann equations for quasiparticle mediated transport, where localised wavepackets are formed from superpositions of single-particle Bloch states. Beyond these bounds, the mean free path of a quasiparticle is of order or smaller than its Compton wavelength, where it is no longer possible to form a coherent quasiparticle from superpositions of Bloch states due to the uncertainty in the single-particle state positions.

Violation of the MIR limit is commonly observed in strongly correlated systems at high temperatures and is often used to suggest that transport in these materials is not described by quasiparticle physics. The ``thermal'' MIR criterion is also a condition on the validity of the Boltzmann description, but is subtly different to the usual MIR criterion as clearly explained by Hartnoll and Mackenzie\cite{Mousatov2020, Hartnoll2021} who refer to it instead as a ``Planckian bound''. Whereas the MIR criterion discerns the ability to form coherent particles from the superposition of Bloch states, the Planckian bound describes the ability of quasiparticles to survive inelastic many-body scattering. 

Despite this, here we find that the Feynman variational method, a quasiparticle theory, predicts mobilities outside of the Planckian bound, in good agreement with diagMC mobility predictions. 

We strongly caution the use of semi-classical mobility theories using Bloch waves as their charge-carrier wavefunction ansatz to model polar materials. 

\subsection{Comparison of the FHIP and diagMC mobilities and a note on dissipation}

In the Feynman variational theory, we see non-monotonic temperature dependence in the mobility. 
At strong coupling there is a range of temperatures where the temperature exponent of the mobility is negative, which begins around $T \simeq \hbar\omega_0$ and ends around some temperature that scales with the Fr\=ohlich coupling parameter $\alpha$. 
The latter high temperature limit marks the transition from strongly coupled polaronic excitations to a thermal electron state (Eq. \ref{eqn:hotmobility}), which is reached asymptotically at large temperatures. 
Compared to the diagMC mobility, we need to go to larger $\alpha$ parameters (beyond 8) to start to see a 'ski jump' rise in mobility with temperature, whereas in diagMC this is seen already at $\alpha = 6$. Though we note that our FHIP results lie within the majority of the Monte-Carlo error bars.

While the temperature-dependence of the FHIP mobility agrees well with the diagMC results, the frequency-dependence differs greatly. 
This has already been investigated\cite{DeFilippis2006} and is due to the harmonic nature of the Feynman trial action. 
The Feynman trial action lacks a dissipative mechanism for the polaron, such that the polaron state described by this model does not lose energy and has an infinite lifetime. 
The spectral function for this model, $A(\Omega) = -2\Im \chi(\Omega)$ (where $\chi$ is the memory function), is a series of delta functions. 
In\cite{DeFilippis2006}, this is corrected by including additional dissipation processes, whose strength is fixed by an exact sum rule. 
This was achieved by directly altering the FHIP memory function, such that the resultant spectral function is a series of Gaussian functions. 
Their resultant frequency-dependent mobility has better agreement with the diagMC mobility. 

Another alternative approach to include dissipation may be to extend the trial Lagrangian in Eqn. (\ref{eqn:triallagrangian}) to incorporate dissipation whilst maintaining that the resulting trial path integral still be evaluable\cite{sels2016, Ichmoukhamedov2022}. 
This would also enable the direct inclusion of anharmonic phonons. 
Applying these generalised trial actions will be the subject of future work.

\subsection{Numerical evaluation of the memory function}

Part of evaluating the FHIP mobility requires a numerical integration in the `memory function' given in Eqn. (\ref{eqn:fhip_chi}). 
While this is usually done by rotating the contour of the integral (given by Eqns. (\ref{eqn:powerseriesimag}) and (\ref{eqn:powerseriesreal})) and expanding as a power-series of special functions, we found that it is far more computationally efficient to directly evaluate the original (non-rotated) integral, using standard adaptive Gauss-Quadrature methods. 
Part of this investigation lead us to derive power-series expansions for the real and imaginary components of the memory function, which we show in the Appendices. The expansion for the imaginary component in terms of Bessel-K functions has been produced before in\cite{Devreese1972}, however we found a new expansion for the real component in terms of Bessel-I and Struve-L functions. 
While we ultimately did not use these expansions in our numeric results presented here, asymptotic evakyatuib of these forms may be useful for future theoretical analysis or numerical calculations.

\subsection{The FHIP initial product state and low-temperature mobility}

In Sec. \ref{Sec:FHIP}, we briefly mentioned that in FHIP\cite{Feynman1962} they assume a nonphysical initial state, which results in an incorrect low-temperature weak-coupling approximation for the dc-mobility with a spurious `$2\beta$' appearing in the denominator of the mobility,
\begin{equation}
    \mu_{\text{FHIP}} = \left( \frac{w}{v} \right)^3 \frac{3e}{2 m^*} \frac{\exp(\beta)}{2\beta \alpha \Omega} \exp\left(\frac{v^2-w^2}{w^2 v}\right).
\end{equation}

This observation is important for understanding the $3/2\beta$ discrepancy between the low temperature FHIP dc-mobility and Kadanoff's dc-mobility\cite{Kadanoff1963} derived from the Boltzmann equation,
\begin{equation}
    \mu_{\text{K}} = \left(\frac{w}{v}\right)^3 \frac{e}{2 m^*} \frac{\exp(\beta)}{\alpha \Omega} \exp\left(\frac{v^2 - w^2}{w^2 v}\right).
\end{equation}

Some have argued that this discrepancy is due to taking the incorrect order of the limits $\Omega \to 0$ and $\alpha \to 0$\cite{Peeters1983mobility}. An alternative form of the low temperature dc-mobility was derived by Los\cite{Los1984, Los2017, Los2018} and Sels\cite{DriesSelsThesis}. Their mobility results differs by a factor of $3$ from Kadonoff and by a factor of $2\beta$ from FHIP,
\begin{equation}
    \mu_{\text{L}} = \left(\frac{w}{v}\right)^3 \frac{3 e}{2 m^*} \frac{\exp(\beta)}{\alpha \Omega} \exp\left(\frac{v^2 - w^2}{w^2 v}\right).
\end{equation}

Sels\cite{DriesSelsThesis} shows that the difference with Kadonoff is because the relaxation time approximation (neglecting the non-vanishing in-scattering term) used by Kadonoff violates particle number conservation, whereas FHIP does not. However, the FHIP approximation relies on a nonphysical initial state for Feynman's polaron model, as mentioned above. 
Further, Los\cite{Los2017, Los2018} shows that not using a factorised initial state of the electron-phonon system results in corrections (although small) due to initial correlations being neglected.

In this work we do not use the low-temperature weak-coupling approximate form of the FHIP mobility, instead we perform a direct numerical integration of the integral in the memory function $\chi$ in Eqn. (\ref{eqn:memoryfunction}).

\subsection{The multimodal extension to the Feynman variation approach}

We compared the free energy and linear response of the polaron evaluated from
the Hellwarth and Biaggio\cite{Hellwarth1999} effective phonon mode method to
our explicit multiple phonon mode method. Applied to the 15 optical solid-state
phonon modes in MAPbI$_3$, we show that our explicit mode method predicts
a slightly higher mobility for temperatures $0$ K to $400$ K, to a maximum of
$20$ \% increase at $100$ K. At $300$ K we predict electron and hole mobilities
of $160$ and $112$ cm$^2$V$^{-1}$s$^{-1}$ respectively. This is to be compared
to our previous predictions of $133$ and $94$ cm$^2$V$^{-1}$s$^{-1}$ for one
effective phonon mode evaluated using Hellwarth and
Biaggio's\cite{Hellwarth1999} `B scheme' (see Eqns. (\ref{eqn:b1}) and
(\ref{eqn:b2})) of $2.25$ THz, as evaluated in our previous work in Ref.
\onlinecite{Frost2017}. 

More importantly, we recover considerable structure in
the complex conductivity and impedance functions as individual phonon modes are
activated. This theory provides a quantitative quantum-mechanical method to
predict the structure we proposed from semi-classical reasoning
in\cite{Leguy2016} (see Figure 10). 
Towards higher temperatures the effective
and explicit methods show the same temperature and frequency dependence - the
quantum details are washed out. 

\subsection{Future work and outlook}

There are many possible extensions of the Feynman polaron approach to increase the accuracy of the approximations, and to more accurately model real systems. 
As discussed, dissipative processes in the trial action would avoid unphysical failures to thermalise and spurious quantum recurrances, most notable in the frequency dependent mobility. 
This requires generalising the trial action. Recently Ichmoukhamedov and Tempere\cite{Ichmoukhamedov2022}, in applying the variational path integral approach to the Bogoliubov-Fr\"ohlich Hamiltonian, extended the trial action to a more general form, and also considered higher-order corrections beyond the Jensen-Feynman inequality. 
While the higher-order corrections are known to be small for the original Fr\"ohlich model\cite{Marshall1970, Lu1992}, they may be important for more general electron-phonon interaction Hamiltonians. 

Recently Houtput et al.\cite{Houtput2021} have extended the Fr\"ohlich model to anharmonic phonon modes. They show that anharmonicity further localises the polaron. As MAPbI$_3$, and other soft polar semiconductors are highly anharmonic, extending the mobility theory of this paper to include anharmonic couplings would be of considerable utility. 

Throughout this paper we have restricted ourselves to a single pair of $v$ and $w$ variational parameters. 
It is possible to generalise the theory to multiple normal modes in the quasi-particle solution, which allows for richer structure in the mobility theory, and a closer approximation to complex multi mode materials. 

The FHIP approach\cite{Feynman1962} is limited to the linear-response regime where the applied field is considered weakly alternating. 
The linear-response regime is sufficient for most technical applications, but non-linear effects may be relevant to interpreting pump-probe THz conductivity measurements. 
The non-linear extensions\cite{Thornber1970, Janssen1995} of FHIP offer a theoretical route to add this in the future.

\section{Acknowledgement}
We thank Andrey Mishchenko for providing their raw Diagrammatic Monte-Carlo data\cite{Mishchenko2019} for co-plotting, and for useful discussions. 
We thank Sergio Ciuchi and the anonymous PRB reviewer for critically reading an earlier version of this manuscript. 
B.A.A.M. is supported by an EPSRC Doctoral Training Award (2446070). 
%
J.M.F. is supported by a Royal Society University Research Fellowship
(URF-R1-191292). 
This work used the Imperial College Research Computing Service\cite{HPC}.  
Via our membership of the UK's HEC Materials Chemistry Consortium, which is funded by EPSRC (EP/R029431), this work used the ARCHER2 UK National Supercomputing Service (http://www.archer2.ac.uk).
Open-source Julia\cite{Julia} codes implementing these methods are available as a repository on
GitHub\cite{GitHub}.



\bibliography{PolaronMobilityBeyondQuasiParticle,PolaronMobilityVibrationalModes}

%
%

\clearpage
\newpage
\onecolumngrid
\appendix

\section{Contour integration of the memory function} 

Following Devreese et al. Ref. \onlinecite{Devreese1972} we derived infinite-power-series expansions of the real and imaginary components of Eqn. (\ref{eqn:fhip_chi}) (or Eqn. (35) in Ref. \onlinecite{Feynman1962}) in terms of Bessel and Struve special functions, and hypergeometric functions. 

The practical computational implementation of these expansions was made difficult by the very high precision required on the special functions to make the expansions converge. Using arbitrary precision numerics, a partially working implementation was developed, but it was discovered that direct numeric integration of Eqn. (\ref{eqn:fhip_chi}) could achieve the same result with less computation time and less complex code.
\newline

\noindent We start by changing the contour of the memory function as done in Ref. \onlinecite{Feynman1962}. The memory function for the polaron is defined to linear order in Ref. \onlinecite{Peeters1984} as $\Sigma(\Omega) = \chi^*(\Omega) / \Omega$, where,
\begin{equation}\label{eqn:memfunc}
    \chi(\Omega) = \int_0^\infty \left[ 1 - e^{i \Omega u} \right] \textrm{Im} S(u)\ du,
\end{equation}
is the 
\begin{equation}
    S(u) = \frac{2\alpha}{3\sqrt{\pi}} \left[D(u) \right]^{-\frac{3}{2}} \left( e^{iu} + \frac{2}{e^{\beta} - 1} \textrm{cos}(u) \right),
\end{equation}
and 
\begin{equation}
    \begin{gathered}
    D(u) = \frac{w^2}{\beta v^2} \left\{a^2 - \beta^2/4 -\ b\ \textrm{cos}(vu)\ \textrm{cosh}(v\beta/2) +\ u^2 -\ i \left[ b\ \textrm{sin}(vu)\ \textrm{sinh}(v\beta/2) + u\beta \right] \right\},
    \end{gathered}
\end{equation}
with $R \equiv (v^2 - w^2) / (w^2 v)$, $a^2 = \beta^2 / 4 + R \beta\ \textrm{coth}(\beta v / 2)$ and $b = R \beta\ /\ \textrm{sinh}(\beta v / 2)$, which are the same as Eqns. (47b) in Ref. \onlinecite{Feynman1962}. 

Solving for the real and imaginary parts of $\Sigma(\Omega)$ gives the real and imaginary parts of $\chi(\Omega)$,
\begin{subequations}
    \begin{align}
        \textrm{Re} \, \chi(\Omega) &= \int_0^\infty \left[ 1 - \textrm{cos}(\Omega u) \right] \textrm{Im} S(u)\ du, \\[1em]
        \textrm{Im} \, \chi(\Omega) &= \int_0^\infty \textrm{sin}(\Omega u)\ \textrm{Im} S(u)\ du .
    \end{align}
\end{subequations}
As both $[1 - \textrm{cos}(\Omega u)]$ and $\textrm{sin}(\Omega)$ are real we can take `Im' outside the integral,
\begin{subequations}
    \begin{align}
        \textrm{Re} \, \chi(\Omega) &= \textrm{Im} \int_0^\infty \left[ 1 - \textrm{cos}(\Omega u) \right] S(u)\ du,  \label{realsus} \\[1em]
        \textrm{Im} \, \chi(\Omega) &= \textrm{Im} \int_0^\infty \textrm{sin}(\Omega u)\ S(u)\ du\ . \label{imagsus}
    \end{align}
\end{subequations}
Now we promote $u \in \mathbb{R}$ to a complex variable $u = x + iy \in \mathbb{C}$. The integrals then become integrals on the complex plane, 
\begin{subequations}
\begin{align}
    \begin{split}
    \textrm{Re} \chi(\Omega) &= \textrm{Im} \int_\Gamma \left[ 1 - \textrm{cos}(\Omega x) \textrm{cosh}(\Omega y) +\ i\ \textrm{sin}(\Omega x) \textrm{sinh}(\Omega y) \right] S(x + iy)\ du,  \label{realsuscomplex} 
    \end{split} \\[1em]
    \begin{split}
    \textrm{Im}\chi(\Omega) &= \textrm{Im} \int_\Gamma \left[ \textrm{sin}(\Omega x)\ \textrm{cosh}(\Omega y) +\ i\ \textrm{cos}(\Omega x)\ \textrm{sinh}(\Omega y) \right] S(x + iy)\ du, \label{imagsuscomplex}
    \end{split}
\end{align}
\end{subequations}
where $\Gamma$ is our contour of integration. To motivate a choice of contour, let's consider the form of $D(x + iy)$ and $S(x + iy)$, 

\begin{equation}
\begin{gathered}
    D(x + i y) = \frac{w^2}{\beta v^2} \left\{\left[ a^2 - \beta^2/4 \right.\right. 
    \left.\left.-\ b\ \textrm{cos}(vx)\ \textrm{cosh}(v( y-\beta/2)) + x^2 + y(\beta - y) \right] \right. \\
    \left.+\ i \left[ b\ \textrm{sin}(vx)\ \textrm{sinh}(v(y-\beta/2)) + 2x(y-\beta/2) \right] \right\}
\end{gathered}
\end{equation}

\begin{equation}
    S(x + i y )
    = \frac{2 \alpha}{3 \sqrt{\pi}} \frac{\textrm{cos}(x + i(y
    - \beta/2))}{\textrm{sinh}(\beta / 2) \left[ D(x + iy) \right]^{\frac{3}{2}}}.
\end{equation}

Now we notice that that $D(x + i y )$ and $S(x + i y )$ are trivially real when $y = \beta / 2$. This gives the results,
\begin{equation}
    D(x + i\beta/2) = \frac{w^2}{\beta v^2} \left[ x^2 + a^2 - b\ \textrm{cos}(vx) \right] \in \mathbb{R}
\end{equation}
\begin{equation}
    \begin{split}
        S(x + i\beta/2) = \frac{2 \alpha}{3 \sqrt{\pi}}
        \frac{\beta^{\frac{3}{2}}}{\textrm{sinh}(\beta / 2)} \left( \frac{v}{w}
        \right)^3 \frac{\textrm{cos}(x)}{\left[x^2 + a^2 - b\ \textrm{cos}(vx)
        \right]^{\frac{3}{2}}} \in \mathbb{R}.
    \end{split}
\end{equation}

\begin{center}
\begin{figure}
\begin{tikzpicture}[decoration={markings,
    mark=at position 2cm   with {\arrow[line width=1pt]{stealth}},
    mark=at position 6cm with {\arrow[line width=1pt]{stealth}},
    mark=at position 9cm   with {\arrow[line width=1pt]{stealth}},
    mark=at position 11cm with {\arrow[line width=1pt]{stealth}},
    mark=at position 14cm   with {\arrow[line width=1pt]{stealth}},
    mark=at position 18cm   with {\arrow[line width=1pt]{stealth}},
    mark=at position 21cm   with {\arrow[line width=1pt]{stealth}},
    mark=at position 23cm   with {\arrow[line width=1pt]{stealth}},
  }]
  \draw[thick, ->] (0,0) -- (9,0) coordinate (xaxis);

  \draw[thick, ->] (0,0) -- (0,5) coordinate (yaxis);

  \node[above] at (xaxis) {$\mathrm{Re}(u)$};

  \node[right]  at (yaxis) {$\mathrm{Im}(u)$};

  \path[draw,blue, line width=0.8pt, postaction=decorate] 
        (0,4)
    --  (8,4)  node[midway, above right, black] {$\Gamma_3$} node[above, black] {$\infty + i\frac{\beta}{2}$}
    --  (8,0)  node[midway, right, black] {$\Gamma_4$} node[below, black] {$\infty$}
    --  (0,0)  node[midway, below, black] {$\Gamma_1$} node[below, black] {$0$} 
    --  (0,4)  node[midway, left, black] {$\Gamma_2$} node[above, left, black] {$i\frac{\beta}{2}$};
\end{tikzpicture}
\caption{\label{fig:integralcontour} The complex contour chosen to transform the integral in Eqn. (\ref{eqn:memfunc}). No singularities lie within the closed contour so the contour integral is zero.}
\end{figure}
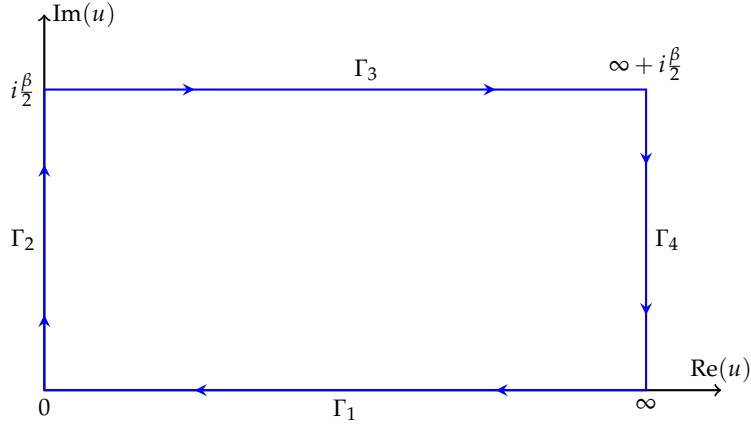
\end{center}

From this, we choose to integrate over the contours $\Gamma_1 \in (\infty + 0i, 0 + 0i] \rightarrow \Gamma_2 \in [0 + i0, 0 + i\beta/2] \rightarrow \Gamma_3 \in [0 + i\beta/2, \infty + i\beta/2) \rightarrow \Gamma_4 \in (\infty + i\beta/2, \infty + 0i)$ as shown in Fig. \ref{fig:integralcontour}. 
Since the integrands in Eqns. (\ref{realsuscomplex}) and (\ref{imagsuscomplex}) are analytic in this region, this closed contour integral will be zero. 
(There is a pole in $\textrm{Im}S(x+iy)$ at $0 + i0$, but this is cancelled by the zero of the elementary/trigonometric functions in front of it at this point.) 
The closing piece of the contour lies at $x \rightarrow \infty$ and can be neglected as $S(x+iy) \rightarrow 0$ in this limit. 

Thus, for the real part of $\chi(\Omega)$ we have,
\begin{equation}
    \begin{aligned}
        \int_0^\infty \left[ 1 - \textrm{cos}(\Omega x) \right] S(x)\ dx &=
        \int_0^{\beta / 2} \left[ 1 - \textrm{cosh}(\Omega y) \right] S(iy)\ d(iy) \\ 
        &+ \int_0^\infty \left[ 1 - \textrm{cos}(\Omega x) \textrm{cosh}\left(\frac{\Omega \beta}{2}\right)\right. 
        \left.+\ i\ \textrm{sin}(\Omega x) \textrm{sinh}\left(\frac{\Omega \beta}{2}\right) \right] S\left(x + \frac{i \beta}{2}\right)\ dx,
    \end{aligned}
\end{equation}
and for the imaginary part of $\chi(\Omega)$ we have,
\begin{equation}
    \begin{aligned}
        \int_0^\infty \textrm{sin}(\Omega x) S(x)\ dx &=
        i\int_0^{\beta / 2} \textrm{sinh}(\Omega y) S(iy)\ d(iy) \\
        &+ \int_0^\infty \left[ \textrm{sin}(\Omega x) \textrm{cosh}\left(\frac{\Omega \beta}{2}\right) +\ i\ \textrm{cos}(\Omega x) \textrm{sinh}\left(\frac{\Omega \beta}{2}\right) \right] S\left(x + \frac{i \beta}{2}\right)\ dx.
    \end{aligned}
\end{equation}

We can now see more clearly why we choose to integrate at $y = \beta / 2$. Since $S(x + i\beta / 2)$ is real, acting `Im' on these integrals will cancel the second integral in the contour integral for $\textrm{Im}\chi(\Omega)$ (which is entirely real), and the third integral for both $\textrm{Re}\chi(\Omega)$ and $\textrm{Im}\chi(\Omega)$ is simplified due to the absence of any cross-terms that would have resulted for other values of $y$ as $S(x + iy)$ would have been complex. To see that the second integral for $\textrm{Im}\chi(\Omega)$ is real, we need to see if $S(iy)$ is real. First, we look at $D(iy)$, which is given by,
\begin{equation}
    \begin{gathered}
        D(iy) = \frac{w^2}{\beta v^2} \left[a^2 - \frac{\beta^2}{4} + y(\beta - y) - b\ \textrm{cosh}\left( vy - \frac{\beta v}{2} \right) \right] \in \mathbb{R},
    \end{gathered}
\end{equation}
and then $S(iy)$ is given by, 
\begin{equation}
    \begin{gathered}
        S(iy) = \frac{2 \alpha}{3 \sqrt{\pi}} \frac{\beta^{3/2}}{\textrm{sinh}(\beta / 2)} \left( \frac{v}{w} \right)^3 \frac{\textrm{cosh}(y - \beta / 2)}{\left[a^2 - \beta^2/4 + y(\beta - y) - b\ \textrm{cosh}(v(y - \beta / 2)) \right]^{3/2}} \in \mathbb{R},
    \end{gathered}
\end{equation}
\newline
so $S(iy)$ is indeed real. Since the second integral for $\textrm{Im}\chi(\Omega)$ has two complex $i$ s and $S(iy)$ is real, the whole integral is entirely real and so it doesn't contribute to $\textrm{Im}\chi(\Omega)$. Unfortunately, $\textrm{Re}\chi(\Omega)$ does not simplify as nicely as $\textrm{Im}\chi(\Omega)$ because the second integral is imaginary and so is still present after taking only the imaginary parts. Nonetheless, for $\textrm{Re}\chi(\Omega)$ we get,

\begin{equation}
    \begin{aligned}
        \textrm{Re}\chi(\Omega) &= \textrm{Im} \int_0^\infty \left[1 - \textrm{cos}(\Omega x)\right] S(x)\ dx \\
        &=\frac{2 \alpha}{3 \sqrt{\pi}} \frac{\beta^{3/2}}{\textrm{sinh}(\beta / 2)} \left( \frac{v}{w} \right)^3 \bigg\{
        \textrm{sinh}\left( \frac{\Omega \beta}{2} \right) \int_0^\infty \frac{\textrm{sin}(\Omega x) \textrm{cos}(x)\ dx}{\left[ x^2 + a^2 - b\ \textrm{cos}(vx) \right]^{3/2}}\\
        &+ \int_0^{\beta/2} \frac{\left[ 1 - \textrm{cosh}(\Omega x) \right] \textrm{cosh}(x - \beta / 2)\ dx }{\left[ a^2 - \beta^2 / 4 + x(\beta - x) - b\ \textrm{cosh}(v(x - \beta / 2)) \right]^{3/2}} \bigg\},
    \end{aligned}
\end{equation}
and for $\textrm{Im}\chi(\Omega)$ we get,
\begin{equation} \label{eqn: ImX}
    \begin{aligned}
         \textrm{Im}\chi(\Omega) = \textrm{Im} \int_0^\infty \textrm{sin}(\Omega x) S(x)\ dx
         = \frac{2 \alpha}{3 \sqrt{\pi}} \frac{\beta^{3/2}\ \textrm{sinh}(\Omega \beta / 2)}{\textrm{sinh}(\beta / 2)} \left( \frac{v}{w} \right)^3 \int_0^\infty \frac{\textrm{cos}(\Omega x) \textrm{cos}(x)\ dx}{\left[x^2 + a^2 - b\ \textrm{cos}(vx) \right]^{3/2}}.
    \end{aligned}
\end{equation}

\newpage
\section{Im$\chi$ expansion in Bessel-K functions} 

In Devreese et al. Ref. \onlinecite{Devreese1972} the integral in Eqn. (\ref{eqn: ImX}) is expanded in an infinite sum of modified Bessel functions of the second-kind. 

Here we follow the same procedure, and arrive at the same result, but provide detailed workings. 

Specifically, we are interested in solving the integral,
\begin{equation} \label{eqn: imx_integral}
    \int_0^\infty \frac{\textrm{cos}(\Omega x) \textrm{cos}(x)\ dx}{\left[x^2 + a^2 - b\ \textrm{cos}(vx) \right]^{3/2}}\ .
\end{equation}
\newline
We start by noticing that,
\begin{equation} \label{eqn: inequality}
    \left| \frac{b\ \textrm{cos}(vx)}{x^2 + a^2} \right| < 1 \quad \textrm{if}\ v > 0\ \textrm{and}\ \beta > 0\ ,
\end{equation}
\newline
so we can do a binomial expansion of the denominator,
\begin{equation}
    \begin{aligned}
        \int_0^\infty \frac{\textrm{cos}(\Omega x) \textrm{cos}(x)}{\left(x^2 + a^2\right)^{3/2}} \left[ 1 - \frac{b\ \textrm{cos}(vx)}{x^2 + a^2}\right]^{-3/2}dx &= \int_0^\infty dx\ \frac{\textrm{cos}(\Omega x) \textrm{cos}(x)}{\left(x^2 + a^2\right)^{3/2}} \sum_{n=0}^\infty \binom{-3/2}{n} \frac{(-b)^n \textrm{cos}^n(vx)}{\left( x^2 + a^2 \right)^n}dx\\
        &= \sum_{n=0}^\infty \binom{-3/2}{n} (-b)^n \int_0^\infty \frac{\textrm{cos}(\Omega x) \textrm{cos}(x) \textrm{cos}^n(vx)}{\left(x^2 + a^2\right)^{n + 3/2}}\ dx,
    \end{aligned}
\end{equation}
\newline
where $\binom{-3/2}{n}$ is a binomial coefficient. Next we expand $\textrm{cos}^n(vx)$ using the power-reduction formula,
\begin{equation}
    \begin{gathered}
        \textrm{cos}^n(vx) = \frac{2}{2^n} \sum_{k=0}^{\floor*{\frac{n-1}{2}}} \binom{n}{k} \textrm{cos}((n -2k)vx) + \frac{(1-n\textrm{mod}2)}{2^{n}} \binom{n}{\frac{n}{2}},
    \end{gathered}
\end{equation}
where the second term comes from even $n$ contributions only. Substituting this into our integral gives,
\begin{equation}
    \begin{gathered}
        \sum_{n=0}^\infty \binom{-3/2}{n} \left(-\frac{b}{2}\right)^n \bigg[
        2 \sum_{k=0}^{\floor*{\frac{n-1}{2}}} \binom{n}{k} \int_0^\infty \frac{\textrm{cos}(\Omega x) \textrm{cos}(x) \textrm{cos}((n -2k)vx)}{\left(x^2 + a^2\right)^{n + 3/2}}\ dx
        + (1-n\textrm{mod}2) \binom{n}{\frac{n}{2}} \int_0^\infty \frac{\textrm{cos}(\Omega x) \textrm{cos}(x)}{\left(x^2 + a^2\right)^{n + 3/2}}\ dx \bigg].
    \end{gathered}
\end{equation}
We can now combine the cosines inside of the integrals into sums of single cosines using,
\begin{equation}
    \begin{aligned}
        \textrm{cos}(\Omega x) \textrm{cos}(x) \textrm{cos}(vx(n - 2k)) &=
        \frac{1}{4} \bigl\{\textrm{cos}(x(\Omega + 1 + v(n-2k)))
        +\textrm{cos}(x(\Omega - 1 + v(n-2k))) \\ 
        &\quad\ \ +\textrm{cos}(x(\Omega + 1 - v(n-2k)))
        +\textrm{cos}(x(\Omega - 1 - v(n-2k))) \bigr\} \\[1em]
        &\equiv \frac{1}{4} \sum_{z_4} \textrm{cos}(xz_{k,4}^n)
    \end{aligned}
\end{equation}
where for brevity we have defined $z_{k,4}^n \in \{\Omega + 1 + v(n-2k),\ \Omega - 1 + v(n-2k),\ \Omega + 1 - v(n-2k),\ \Omega - 1 - v(n-2k) \}$. Likewise,
\begin{equation}
    \begin{gathered}
        \textrm{cos}(\Omega x) \textrm{cos}(x) =\frac{1}{2} \bigl\{\textrm{cos}(x(\Omega + 1)) +\ \textrm{cos}(x(\Omega - 1))\bigr\} \\[1em]
        \equiv \frac{1}{2} \sum_{z_2} \textrm{cos}(xz_2)
    \end{gathered}
\end{equation}
where for brevity we have defined $z_2 \in \{\Omega + 1,\ \Omega - 1\}$. Substituting these into our expansion gives,
\begin{equation}
    \begin{aligned}
        \sum_{n=0}^\infty \binom{-3/2}{n} \left(-\frac{b}{2}\right)^n \bigg[2\sum_{k=0}^{\floor*{\frac{n-1}{2}}} \binom{n}{k} \sum_{z_4} \int_0^\infty \frac{\textrm{cos}(xz^n_{k,4}(\Omega))}{\left(x^2 + a^2\right)^{n + 3/2}}\ dx
        + (1-n\textrm{mod}2) \binom{n}{\frac{n}{2}} \sum_{z_2} \int_0^\infty \frac{ \textrm{cos}(xz_2(\Omega))}{\left(x^2 + a^2\right)^{n + 3/2}}\ dx \bigg].
    \end{aligned}
\end{equation}
We now have a lot of integrals of the form,
\begin{equation} \label{eqn: cosintegral}
    \int_0^\infty \frac{\textrm{cos}(xz)}{\left(x^2 + a^2\right)^{n + 3/2}}\ dx,
\end{equation}
which is an integral representation for modified Bessel functions of the second kind,
\begin{equation}
    \begin{aligned}
        \int_0^\infty \frac{\textrm{cos}(xz)\ dx}{\left(x^2 + a^2\right)^{n + 3/2}} &= \frac{\sqrt{\pi}}{\Gamma(n + 3/2)} K_{n+1}(|z|a) \bigg|\frac{z}{2a}\bigg|^{n + 1} \\
        &\equiv B_n(z)
    \end{aligned}
\end{equation}
Thus, overall we can expand Im$\chi(\Omega)$ in a series of these bessel functions,
\begin{equation}
    \begin{aligned}
        \textrm{Im}\chi(\Omega) = \frac{2\alpha \beta^{\frac{3}{2}}}{3\sqrt{\pi}} \frac{\textrm{sinh}(\frac{\Omega \beta}{2})}{\textrm{sinh}(\frac{\beta}{2})} \left(  \frac{v}{w}\right)^3 \sum_{n=0}^\infty \binom{-\frac{3}{2}}{n} \left(-\frac{b}{2}\right)^n \bigg[\sum_{k=0}^{\floor*{\frac{n-1}{2}}} \binom{n}{k} &\sum_{z_4} B_{n}(z^n_{k,4}(\Omega)) + (1-n\textrm{mod}2) \binom{n}{\frac{n}{2}} &\sum_{z_2} B_{n}(z_2(\Omega))\bigg]
    \end{aligned}
\end{equation}
where $a^2 = \beta^2 / 4 + R \beta\ \textrm{coth}(\beta v / 2)$, $b = R \beta\ /\ \textrm{sinh}(\beta v / 2)$ and $R = (v^2 - w^2) / (w^2 v)$. Also, $z_{k,4}^n(\Omega) \in \{\Omega + 1 + v(n-2k),\ \Omega - 1 + v(n-2k),\ \Omega + 1 - v(n-2k),\ \Omega - 1 - v(n-2k) \}$ and $z_2(\Omega) \in \{\Omega + 1,\ \Omega - 1\}$.\\

\newpage
\section{Re$\chi$ expansion in Bessel-I, Struve-L and $_1F_2$ hypergeometric functions} 

Motivated by the expansion of Im$\chi(\Omega)$ in Devreese et al.\cite{Devreese1972} we provide a similar expansion for Re$\chi(\Omega)$. 

We follow a similar procedure as for Im$\chi(\Omega)$ and notice that our efforts focus on solving the integrals,
\begin{equation}
    \int_0^\infty \frac{\textrm{sin}(\Omega x) \textrm{cos}(x)\ dx}{\left[ x^2 + a^2 - b\ \textrm{cos}(vx) \right]^{3/2}},
\end{equation}
\begin{equation}
    \int_0^{\beta/2} \frac{\left[ 1 - \textrm{cosh}(\Omega x) \right] \textrm{cosh}(x - \beta / 2)\ dx }{\left[ a^2 - \beta^2 / 4 + x(\beta - x) - b\ \textrm{cosh}(v(x - \beta / 2)) \right]^{3/2}}.
\end{equation}
The first integral is very similar to Eqn. (\ref{eqn: imx_integral}), just with a cosine swapped out for a sine. Following a similar procedure as for Eqn. (\ref{eqn: imx_integral}) gives,
\begin{equation}
    \begin{gathered}
         \int_0^\infty \frac{\textrm{sin}(\Omega x) \textrm{cos}(x)\ dx}{\left[ x^2 + a^2 - b\ \textrm{cos}(vx) \right]^{3/2}} = \\
         \sum_{n=0}^\infty \binom{-3/2}{n} \left(-\frac{b}{2}\right)^n \bigg[2\sum_{k=0}^{\floor*{\frac{n-1}{2}}} \binom{n}{k} \sum_{z_4} \int_0^\infty \frac{\textrm{sin}(xz^n_{k,4}(\Omega))}{\left(x^2 + a^2\right)^{n + 3/2}}\ dx
         +(1-n\textrm{mod}2) \binom{n}{\frac{n}{2}} \sum_{z_2} \int_0^\infty \frac{ \textrm{sin}(xz_2(\Omega))}{\left(x^2 + a^2\right)^{n + 3/2}}\ dx \bigg],
    \end{gathered}
\end{equation}
where we now look for any special functions for which,
\begin{equation}
    \int_0^\infty \frac{\textrm{sin}(xz)}{\left(x^2 + a^2\right)^{n + 3/2}}\ dx
\end{equation}
is the integral representation. We found that,
\begin{equation}
    \begin{aligned}
        \int_0^\infty \frac{\textrm{sin}(xz)}{\left(x^2 + a^2\right)^{n + 3/2}}\ dx &= \frac{\sqrt{\pi}}{2}\frac{\Gamma(-\frac{1}{2}-n)\  \textrm{sgn}(z)|z|^{n+1}}{(2a)^{n+1}}\left[ I_{n+1}(|z|a) - \textbf{L}_{-(n+1)}(|z|a) \right] \\
        &\equiv J_n(z)
    \end{aligned}
\end{equation}
for $n \geq 0$ and $a \geq 0$. Here $\textrm{sgn(x)}$ is the signum function, $I_{n}(x)$ is the modified Bessel function of the first kind, $\textbf{L}_{n}(x)$ is the modified Struve function. Therefore, for Re$\chi(\Omega)$ we have,
\begin{equation}
    \begin{aligned}
        \textrm{Re}\chi(\Omega) &=
        \frac{2 \alpha \beta^{3/2}}{3 \sqrt{\pi}} \frac{\sinh\left( \frac{\Omega \beta}{2} \right)}{\sinh(\beta / 2)} \left( \frac{v}{w} \right)^3 \Bigg\{ \sum_{n = 0}^\infty \binom{-\frac{3}{2}}{n} \left(\frac{b}{2}\right)^n \bigg[ \sum_{k=0}^{\floor*{\frac{n-1}{2}}} \binom{n}{k} \sum_{z_4} J_{n}(z_{k,4}^n(\Omega)) + (1-n\textrm{mod}2) \binom{n}{\frac{n}{2}} \sum_{z_2} J_{n}(z_2(\Omega)) \bigg]\Bigg\} \\
        &+ \frac{2 \alpha}{3 \sqrt{\pi}} \frac{\beta^{3/2}}{\sinh(\beta / 2)} \left( \frac{v}{w} \right)^3\int_0^{\beta/2} \frac{\left[ 1 - \cosh(\Omega x) \right] \cosh(x - \beta / 2)\ dx }{\left[ a^2 - \beta^2 / 4 + x(\beta - x) - b\ \cosh(v(x - \beta / 2)) \right]^{3/2}} 
    \end{aligned}
\end{equation}
where $a$, $b$, $z_4$ and $z_2$ are the same as before.

To expand the second integral with the hyperbolic integrand is more complicated. We start by doing a change of variables $x \rightarrow (1 - x) \beta / 2$ to transform the denominator into a similar form as before and to change the limits to $[0, 1]$,

\begin{equation}
    \int_0^{\beta/2} \frac{\left[ 1 - \cosh(\Omega x) \right] \cosh(x - \beta / 2)\ dx }{\left[ a^2 - \beta^2 / 4 + x(\beta - x) - b\ \cosh(v(x - \beta / 2)) \right]^{3/2}} \longrightarrow \frac{\beta}{2} \int_0^1 \frac{[1 - \cosh(\Omega\beta [1 - x] / 2)] \cosh(\beta x / 2) dx}{[a^2 - (\beta x / 2)^2 - b \cosh(\beta v x / 2)]^{3/2}}.
\end{equation}
Now we see that for $x \in [0, 1]$
\begin{equation}
    \bigg| \frac{b\cosh(v\beta x/2)}{a^2 - (\beta x / 2)^2} \bigg| < 1 \quad \textrm{if } v > 0 \textrm{ and } \beta > 0
\end{equation}
so we can do a binomial expansion of the denominator as before,
\begin{equation}
    \begin{aligned}
        \sum_{n = 0}^\infty \binom{-\frac{3}{2}}{n} \left( \frac{2}{\beta}\right)^{2n+2} (-b)^n &\int_0^1 \frac{[1 - \cosh(\Omega\beta[1-x]/2)]\cosh(\beta x/2)\cosh^n(v\beta x / 2)}{((2a/\beta)^2 - x^2)^{n+3/2}} dx.
    \end{aligned}
\end{equation}
Then we do another binomial expansion of the remaining denominator
\begin{equation}
    \begin{aligned}
        \sum_{n = 0}^\infty \binom{-\frac{3}{2}}{n} \left( \frac{2}{\beta}\right)^{2n+2} (-b)^n \sum_{m = 0}^\infty &\binom{-n-\frac{3}{2}}{m} (-1)^m \left( \frac{\beta}{2a}\right)^{2n + 2m + 3} \\
        &\times \int_0^1 \left[1 - \cosh\left(\frac{\Omega\beta[1-x]}{2}\right)\right]\cosh\left(\frac{\beta x}{2}\right)\cosh^n\left(\frac{v\beta x}{2}\right) x^{2m} dx.
    \end{aligned}
\end{equation}
We can then expand the product of hyperbolic cosines in the integrand,
\begin{equation}\label{eqn:bla}
    \begin{aligned}
        &\sum_{n = 0}^\infty \binom{-\frac{3}{2}}{n} \left( \frac{2}{\beta}\right)^{2n+2} (-b)^n \sum_{m = 0}^\infty \binom{-n-\frac{3}{2}}{m} (-1)^m \left( \frac{\beta}{2a}\right)^{2n + 2m + 3} \frac{1}{2^n}  \\
        &\times \Bigg\{\binom{n}{\frac{n}{2}}(1 - n \textrm{mod} 2) \left[ \int_0^1 \frac{\cosh(\frac{\beta z_1 x}{2})}{x^{-2m}} dx - \frac{1}{2} \sum_{z_2} \left( \cosh\left(\frac{\Omega\beta}{2}\right) \int_0^1 \frac{\cosh(\frac{\beta z_2 x }{2})}{x^{-2m}} dx - \sinh\left(\frac{\Omega\beta}{2}\right) \int_0^1 \frac{\sinh(\frac{\beta z_2 x}{2})}{x^{-2m}} dx \right) \right] \\
        &+ \sum_{k=0}^{\floor*{\frac{n-1}{2}}} \binom{n}{k} \left[ \sum_{z_3} \int_0^1 \frac{\cosh(\frac{\Omega\beta z_{3} x}{2})}{x^{-2m}}dx - \frac{1}{2} \sum_{z_4} \left( \cosh\left(\frac{\Omega\beta}{2}\right) \int^1_0 \frac{\cosh(\frac{\Omega\beta z_{4} x}{2})}{x^{-2m}} dx - \sinh\left(\frac{\Omega\beta}{2}\right) \int^1_0 \frac{\sinh(\frac{\Omega\beta z_{4} x}{2})}{x^{-2m}} dx \right) \right] \Bigg\}
    \end{aligned}
\end{equation}
where $z_1 = 1$, $z_2(\Omega) \in \{ \Omega + 1,\ \Omega - 1 \}$, $z_{k,3}^n \in \{ 1 + v(n - 2k),\ 1 - v(n - 2k) \}$ and $z_{k,4}^n(\Omega) \in \{ \Omega + 1 + v(n - 2k),\ \Omega - 1 + v(n - 2k),\ \Omega + 1 - v(n - 2k),\ \Omega - 1 - v(n - 2k) \}$. 
\newline
Now we have two integrals of the forms
\begin{equation}
    \int_0^1 \cosh(zx) x^{2m} dx, \qquad \int^1_0\sinh(zx) x^{2m} dx,
\end{equation}
which are the integral forms of the generalised hypergeometric functions
\begin{subequations}
\begin{equation}
    \int_0^1 \cosh(zx) x^{2m} dx = \pFq{1}{2}{m + \frac{1}{2}}{\frac{1}{2}, m + \frac{3}{2}}{\frac{z^2}{4}} = \sum_{t = 0}^\infty \frac{z^{2t}}{(2t+2m+1)(2t)!}, \quad m > -\frac{1}{2}
\end{equation}
\begin{equation}
    \int_0^1 \sinh(zx) x^{2m} dx = \frac{z}{2m + 2} \pFq{1}{2}{m + 1}{\frac{3}{2}, m + 2}{\frac{z^2}{4}} = \sum_{t=0}^\infty \frac{z^{2t+1}}{(2t+2m+2)(2t+1)!}, \quad m > -1.
\end{equation}
\end{subequations}
For brevity, we will define
\begin{subequations}
\begin{equation}
    _1F_2^c(z) \equiv \pFq{1}{2}{m + \frac{1}{2}}{\frac{1}{2}, m + \frac{3}{2}}{\frac{\beta^2z^2}{16}} = \sum_{t = 0}^\infty \frac{(\beta z / 2)^{2t}}{(2t+2m+1)(2t)!}
\end{equation}
\begin{equation}
    _1F_2^s(z) \equiv \frac{\beta z}{4m + 4} \pFq{1}{2}{m + 1}{\frac{3}{2}, m + 2}{\frac{\beta^2 z^2}{16}} = \sum_{t=0}^\infty \frac{(\beta z / 2)^{2t+1}}{(2t+2m+2)(2t+1)!}
\end{equation}
\end{subequations}
so that Eqn. (\ref{eqn:bla}) becomes
\begin{equation}
    \begin{aligned}
        &\sum_{n = 0}^\infty \binom{-\frac{3}{2}}{n} \left( \frac{2}{\beta}\right)^{2n+2} (-b)^n \sum_{m = 0}^\infty \binom{-n-\frac{3}{2}}{m} (-1)^m \left( \frac{\beta}{2a}\right)^{2n + 2m + 3} \frac{1}{2^n}  \\
        &\times \Bigg\{\binom{n}{\frac{n}{2}}(1 - n \textrm{mod} 2) \left[ _1F_2^c(z_1) - \frac{1}{2} \sum_{z_2} \left( \cosh\left(\frac{\Omega\beta}{2}\right)\ _1F_2^c(z_2) - \sinh\left(\frac{\Omega\beta}{2}\right)\ _1F_2^s(z_2) \right) \right] \\
        &+ \sum_{k=0}^{\floor*{\frac{n-1}{2}}} \binom{n}{k} \left[ \sum_{z_3}\ _1F_2^c(z_3) - \frac{1}{2} \sum_{z_4} \left( \cosh\left(\frac{\Omega\beta}{2}\right)\ _1F_2^c(z_4) - \sinh\left(\frac{\Omega\beta}{2}\right)\ _1F_2^s(z_4) \right) \right] \Bigg\}
    \end{aligned}
\end{equation}
which we can reduce further by defining
\begin{equation}
    M^{c/s}_{n}(z) \equiv \sum_{m = 0}^\infty \binom{-n-\frac{3}{2}}{m} (-1)^m a^{-2(n+m+1)} \left(\frac{\beta}{2}\right)^{2m + 1}\ _1F_2^{c/s}(z) 
\end{equation}
to give
\begin{equation}
    \begin{aligned}
        \sum_{n = 0}^\infty \binom{-\frac{3}{2}}{n} \left(\frac{-b}{2}\right)^n \Bigg\{ \binom{n}{\frac{n}{2}}(1 - n \textrm{mod} 2) \Biggl[ M^{c}_n(z_1) - &\frac{1}{2} \sum_{z_2} \left( \cosh\left(\frac{\Omega\beta}{2}\right) M^{c}_n(z_2) - \sinh\left(\frac{\Omega\beta}{2}\right) M^{s}_n(z_2) \right) \Biggr] \\
        + \sum_{k=0}^{\floor*{\frac{n-1}{2}}} \binom{n}{k} \Biggl[ \sum_{z_3} M^{c}_n(z_3) - &\frac{1}{2} \sum_{z_4} \left( \cosh\left(\frac{\Omega\beta}{2}\right) M^{c}_n(z_4) - \sinh\left(\frac{\Omega\beta}{2}\right) M^{s}_n(z_4) \right) \Biggr] \Bigg\}.
    \end{aligned}
\end{equation}
Combining this with the rest of $\textrm{Re}\chi(\Omega)$ gives
\begin{equation}
    \begin{aligned}
        \textrm{Re}\chi(\Omega) = \frac{2\alpha \beta^{3/2} v^3}{3\sqrt{\pi} w^3 \sinh(\beta/2)} \sum_{n=0}^\infty \binom{-\frac{3}{2}}{n} \left(-\frac{b}{2}\right)^n \qquad\qquad\qquad\qquad\qquad\qquad\qquad\qquad\qquad\qquad\qquad\qquad& \\ 
        \times \Bigg\{ \binom{n}{\frac{n}{2}} (1 - n\textrm{mod}2) \left[ M^{c}_n(1) + \frac{1}{2} \sum_{z_2} \left( \sinh\left(\frac{\Omega\beta}{2}\right) \left[ M^{s}_n(z_2(\Omega)) + J_n(z_2(\Omega)) \right] - \cosh\left(\frac{\Omega\beta}{2}\right) M^{c}_n(z_2(\Omega)) \right) \right]&\\
        + \sum_{k = 0}^{\floor*{\frac{n-1}{2}}} \binom{n}{k} \left[ \sum_{z_3} M^{c}_n(z_{k,3}^n) + \frac{1}{2} \sum_{z_4} \left( \sinh\left(\frac{\Omega\beta}{2}\right) \left[ M^{s}_n(z_{k,4}^n(\Omega)) + J_n(z_{k,4}^n(\Omega)) \right] - \cosh\left(\frac{\Omega\beta}{2}\right) M^{s}_n(z^n_{k,4}(\Omega)) \right) \right]&\Bigg\}.
    \end{aligned}
\end{equation}
So, altogether we have the expansion for the memory function
\begin{equation}
    \begin{aligned}
        \chi(\Omega) = \frac{2\alpha \beta^{3/2} v^3}{3\sqrt{\pi} w^3 \sinh(\beta/2)} \sum_{n=0}^\infty \binom{-\frac{3}{2}}{n} \left(-\frac{b}{2}\right)^n \qquad\qquad\qquad\qquad\qquad\qquad\qquad\qquad\qquad\qquad\qquad\qquad& \\ 
        \times \Bigg\{ \binom{n}{\frac{n}{2}} (1 - n\textrm{mod}2) \left[ M^{c}_n(1) + \frac{1}{2} \sum_{z_2} \left( \sinh\left(\frac{\Omega\beta}{2}\right) \left[ M^{s}_n(z_2) + J_n(z_2) + iB_n(z_2) \right] - \cosh\left(\frac{\Omega\beta}{2}\right) M^{c}_n(z_2(\Omega)) \right) \right]&\\
        + \sum_{k = 0}^{\floor*{\frac{n-1}{2}}} \binom{n}{k} \left[ \sum_{z_3} M^{c}_n(z_{k,3}^n) + \frac{1}{2} \sum_{z_4} \left( \sinh\left(\frac{\Omega\beta}{2}\right) \left[ M^{s}_n(z_{k,4}^n) + J_n(z_{k,4}^n) + iB_n(z_{k,4}^n)) \right] - \cosh\left(\frac{\Omega\beta}{2}\right) M^{s}_n(z^n_{k,4}) \right) \right]&\Bigg\}.
    \end{aligned}
\end{equation}
\end{document}